\documentclass[
 floatfix,
 reprint,
 aps,
]{revtex4-2}

\usepackage{amsmath}
\usepackage{amssymb}
\usepackage{easyReview}
\usepackage{graphicx}
\usepackage{dcolumn}
\usepackage{bm}
\usepackage{xcolor}
\usepackage{braket}

\begin{document}

\preprint{APS/123-QED}

\title{Two-photon quantum state tomography of photonic qubits}

\author{Guilherme P. Temporão}
\email{temporao@puc-rio.br}
\author{Pedro Ripper}
\email{pripper@opto.cetuc.puc-rio.br}
 \altaffiliation[also at ]{PSR, Brazil}
 \affiliation{Department of Electrical Engineering, Pontifical Catholic University of Rio de Janeiro, Brazil}
\author{Thiago B. Guerreiro}%
 \email{barbosa@puc-rio.br}
\affiliation{Department of Physics, Pontifical Catholic University of Rio de Janeiro, Brazil}
\author{Gustavo C. do Amaral}%
 \email{gustavo.castrodoamaral@tno.nl}
\affiliation{Quantum Technology Department, The Netherlands Organization for Applied Scientific Research, TNO, The Netherlands}

\date{\today}%

\begin{abstract}
We provide a tool for measuring the Stokes parameters and the degree of polarization of single photons by employing second order interference, namely the Hong-Ou-Mandel (HOM) interferometer. It is shown that the technique is able to distinguish a partially polarized photon where the polarization state is coupled to an internal degree of freedom, such as time of arrival, from partial polarization due to external entanglement with the environment. The method does not directly resort to any kind of polarization-selective components and therefore is not limited by the extinction ratio of polarizers. Moreover, the technique can be generalized to any two-level encoding of quantum information in single photons, such as time-bin or orbital angular momentum qubits.   

\end{abstract}

\maketitle

\section{\label{sec:intro}Introduction}

The need for single-photon sources (SPS) is increasing for many applications, such as quantum communication, photonic quantum computing and quantum metrology~\cite{Rob2007,Couteau2023,OBrien07}. Therefore, tools for characterization of such sources are of paramount importance. One of the most important aspects of a SPS is the polarization purity of the generated photons, usually called degree of polarization (DOP), which can be measured by standard polarimetry, i.e., polarization Quantum State Tomography (QST)~\cite{Altepeter05,Resch05,Takesue09,Titchener18,Peters2003}. In standard QST, the polarization state of the photons is determined by a succession of measurements of Pauli operators, which are measurements acting on individual photons, one at a time. However, polarization state measurements employing two-photon interference, namely the Hong-Ou-Mandel (HOM) effect~\cite{HOM_original}, have been proposed for direct measurement of the polarization purity of photons, without the need for QST \cite{steinberg}. In fact, a previous result under the context of quantum circuits had already shown that the purity - or any other functional - of a quantum state can be directly measured by performing a Swap Test (ST) on two copies of it \cite{Ekert2002}, where the ST is an operation known to be equivalent to the HOM effect \cite{quantumfp}.  Recently, HOM interference has also been employed for tracking changes in the polarization state over time, with many advantages over the standard methods~\cite{photonics10010072, Harnchaiwat:20,cortes2022sample}.

In this work, we use the HOM effect for a complete characterization of the Stokes parameters and DOP of single photons, solely relying on the second-order interference between two photons from a SPS, without the need for generation of any reference state or deploying any kind of polarization-selective component. This method enables the characterization of additional degrees of freedom of the photons, even those which the detectors are unable to resolve on their own. Moreover, this characterization is performed with a significantly smaller number of measurements if compared to classical measurements, which can be regarded as a kind of quantum advantage~\cite{doi:10.1126/science.abn7293}. As will be shown in calculations and simulations in quantum circuits, the proposed method can detect decoherence effects due to the coupling of the polarization state of the photon to other internal degrees of freedom, such as time of arrival, spatial mode or frequency mode, and therefore distinguishing them from external couplings to the environment or statistical mixtures. The method has applications beyond polarimetry, as it can be generalized to characterize any qubit encoding in a single photon. 
Towards that end, we compare the proposed method against standard QST on a quantum computer simulator to demonstrate its expected performance and encoding generality.

\section{\label{sec:method}Method}

\begin{figure}[b!]
    \centering
    \includegraphics[width=0.35\textwidth]{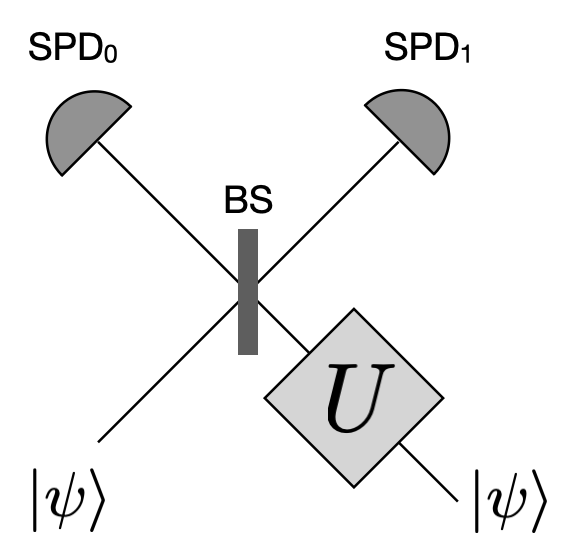}
    \caption{Scheme for determining the polarization state of single photons. Two identical photons represented by the quantum state $\ket{\psi}$, where one of them is subjected to a unitary operator $U$, interfere in a beamsplitter (BS) and the probability of coincidence counts in the single-photon detectors (SPD) is determined, according to Eq.~\ref{eq:one}. If $U$ is replaced successively by the Pauli operators $\sigma_j$, the coincidence probabilities $P_{coinc}(\sigma_j)$ can be used to calculate the DOP and the absolute value of the Stokes parameters of the original photons. Additional steps may be added for obtaining full information on the quantum state, as described in Appendix \ref{app:a}.}
    \label{fig:one}
\end{figure}

Fig.~\ref{fig:one} shows the basic idea of the method, which consists of interfering two single photons on a beamsplitter (BS). Here, the source producing the two input single-photon states are not necessarily the same, the only requirement being that the quantum state originally encoded onto the two photons (denoted $\ket{\psi}$ in Fig.~\ref{fig:one}) is the same; furthermore, before reaching the beamsplitter, one of the two input states is subjected to a polarization transformation represented by the unitary operation $U$. Let $\mathcal{H}$ be the Hilbert space describing all degrees of freedom of the photons, and write $\mathcal{H} = \mathcal{H}_{\text{pol}}\otimes\mathcal{H}_\text{o}$, where $\mathcal{H}_\text{pol}$ is the (two-dimensional) Hilbert space associated with the polarization state of the photons and $\mathcal{H}_\text{o}$ is the Hilbert space associated with the remaining degrees of freedom. Whenever the incoming states are pure (i.e., there is no entanglement between the photons and any external degree of freedom), we can write a generic input state as:
\begin{equation}
\ket{\psi} = \sqrt{p}\ket{\phi}\ket{\xi}+e^{i\theta}\sqrt{1-p}\ket{\phi^\perp}\ket{\xi^\perp}
\label{eq:one1}
\end{equation}
where $\ket{\phi} \in \mathcal{H}_\text{pol}$, $\ket{\xi} \in \mathcal{H}_\text{o}$, $\theta$ is a generic phase, the sign $\perp$ means a orthogonal state and $p$ is a real parameter $(0 \leq p \leq 1)$ that can be interpreted as a probability of finding the photon in the polarization state $\ket{\phi}$. Note that Eq.~\ref{eq:one1} describes a partially polarized photon whenever $0 < p < 1$. 

Using two threshold single-photon detectors (SPD) in the BS outputs, the probability of obtaining a coincidence count $P_{coinc}$ as a function of the applied unitary operator $U$ is given by the following expression:
\begin{equation}
P_{coinc}(U) = \frac{1-\mathcal{F}(U)}{2}
\label{eq:one}
\end{equation}
where $\mathcal{F}(U)$ is the $\textit{fidelity}$ between the two input quantum states impinging on the BS when an unitary operator $U \in \mathcal{B}[\mathcal{H}_\text{pol}]$ is applied to one of the inputs. The derivation of Eq. \ref{eq:one} can be found in Appendix \ref{app:b}. Note that, due to the single-photon nature of the input states, the HOM visibility reaches 100\% when the photons are indistinguishable; this, however, does not limit the effectiveness of the method when different photon statistics are considered, but it does impact its performance~\cite{amaral2019characterization}. The fidelity is given by:
\begin{equation}
\mathcal{F}(U) = |\braket{\psi | U\otimes I | \psi}|^2
\label{eq:two2}
\end{equation}
where $I$ is the identity operator acting on $\mathcal{H}_\text{o}$.

Now recall from standard polarimetry that the (normalized) Stokes parameters are defined, for pure polarization states, as \cite{Berglund2000}:
\begin{equation}
s_j \equiv \text{tr}\left(\ket{\psi}\bra{\psi}\sigma_j\right) = \braket{\phi | \sigma_j | \phi}
\label{eq:new}
\end{equation}
i.e., they correspond to the expectation values of the Pauli spin operators. Under the context of polarization states, they can be written as: 
\begin{eqnarray}
\sigma_1 \equiv && \ket{H}\bra{H} - \ket{V}\bra{V} \nonumber \\
\sigma_2 \equiv && \ket{+45}\bra{+45} - \ket{-45}\bra{-45}\nonumber \\
\sigma_3 \equiv && \ket{RC}\bra{RC} - \ket{LC}\bra{LC}
\label{eq:three}
\end{eqnarray}
where $\{\ket{H},\ket{V}\}$ are the horizontal and vertical polarizations, $\{\ket{+45},\ket{-45}\}$ are the diagonal/anti-diagonal polarizations and $\{\ket{RC},\ket{LC}\}$ correspond to the right-circular/left-circular pure polarization states. The reader should be aware of the unusual labeling of the Pauli operators; this is due to the re-labeling of the axes when going from the Bloch sphere to Poincaré sphere.

Let us now consider a simplified version of the protocol, where we are only interested in finding the DOP of the photons; we also assume for the moment that there is no external entanglement taking place - see section \ref{sec:ext_ent} for a discussion and Appendix \ref{app:a} for the full protocol. The idea consists in sequentially selecting $U = \sigma_j$, for $j = 1,2,3$. It is straightforward from Eq.~\ref{eq:two2} and Eq.~\ref{eq:one1} that
\begin{eqnarray}
\mathcal{F}(\sigma_j) && = |p\braket{\phi | \sigma_j | \phi} + (1-p)\braket{\phi^\perp | \sigma_j | \phi^\perp}|^2 \nonumber \\ && = [(2p-1)s_j]^2 \nonumber \\ && = \langle s_j \rangle^2 
\label{eq:four4}
\end{eqnarray}
where $\langle s_j \rangle$ are defined as the time averaged Stokes parameters. The probabilities of coincidence as a function of the applied Pauli operator can be found by replacing Eq.~\ref{eq:four4} on \ref{eq:one}:
\begin{equation}
P_{coinc}(\sigma_j) = \frac{1-\langle s_j \rangle^2}{2}
\label{eq:four}
\end{equation}

The degree of polarization (DOP), on the other hand, corresponds to the norm of the Stokes vector. Using the usual definition for DOP, we have:
\begin{equation}
DOP = \left[\sum_{j=1}^3\langle s_j \rangle^2\right]^{1/2}
\label{eq:five}
\end{equation}

Equation \ref{eq:five} is somewhat surprising, as the degree of polarization of light can be determined without any polarization-selective component, such as polarizers or polarizing beamsplitters (PBS). It is straightforward to show that the state $\ket{\psi}$ given by Eq.~\ref{eq:one1} has $DOP = |2p-1|$.

Please note that the DOP of Eq. \ref{eq:five} is unrelated to the purity of the photons. Previous works employed HOM interference \cite{steinberg} or, more generally, the Swap Test \cite{Ekert2002} for directly determining the purity $v \equiv \text{tr}(\rho)^2$ of an incoming photon described by its density operator $\rho$. The DOP, on the other hand, can be understood as an equivalent measure of purity for the {\it reduced density operator} that describes only the polarization degree of freedom, not the global state $\rho$. The main advantage of our method is that the polarization reduced density operator can be reconstructed without any generation of reference states, which would require local sources identical to the SPS under test (except for the polarization degree of freedom) and polarization-sensitive components for preparation of the polarization (pure) states, therefore rendering the process completely unfeasible.  In this work, the purity $v$ corresponds to the coincidence probability measured when $U = I$, which will also be employed in the discrimination between internal and external entanglement in section \ref{sec:III}.

It should be stressed that the presented method so far is able to determine the absolute values of the Stokes parameters, $|s_j|$, but not their signs. Additional measurements, and the inclusion of a polarization dependent loss (PDL) element in front of one of the detectors, is a possible way to circumvent this limitation; the full protocol for complete characterization of the polarization state is described in Appendix \ref{app:a}.

\section{Application of the method in partially polarized light}
\label{sec:III}

Partially polarized light may be obtained in two ways. The first one is by incoherently mixing polarized photons, whereas the second
one consists of, for each photon, entangling the polarization state
with another degree of freedom, which can belong to the same photon or to some external environment, thus obtaining a
partially polarized photon \cite{Jeffrey2004,Peters2005}. In the next sections, we discuss the cases of ``internal" and ``external" entanglement in partially polarized photons, which correspond, respectively, to internal degrees of freedom of the photon and external quantum systems, commonly referred to as the environment. The case of incoherent mixtures, as will be discussed, is indistinguishable from external entanglement. 

\subsection{\label{sec:int_ent}``Internal" entanglement}

We provide now an example where the polarization degree of freedom is entangled with the photon time-of-arrival, usually called ``time-bin" in the context of quantum communications~\cite{Ivan2003}, in order to show that the definition in Eq.~\ref{eq:five} coincides with the usual DOP definition for such states. Let $\{\ket{t_0},\ket{t_1}\}$ represent the states where the photon wave-packets are centered at times of arrival $t_0$ and $t_1 = t_0 + \Delta t$, where the delay $\Delta t$ between each bin is longer than the photon coherence time $\tau_c$.

\begin{figure}[t!]
    \centering
    \includegraphics[width=\linewidth]{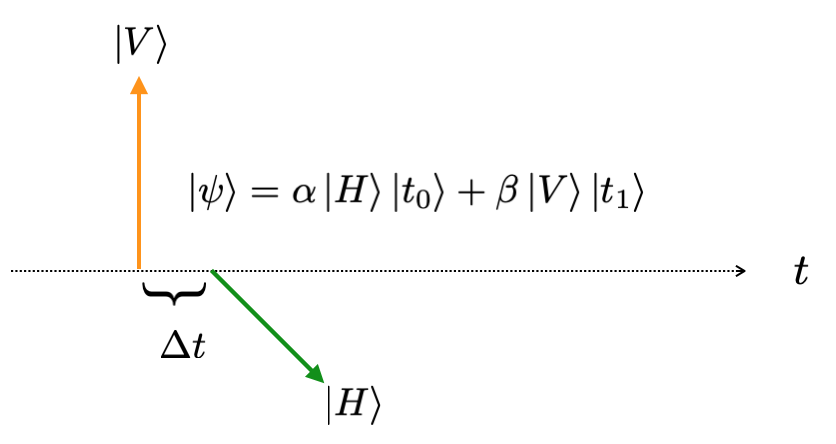}
    \caption{Partially polarized light due to polarization-time coupling, where $\alpha,\beta$ are complex numbers such that $|\alpha|^2 + |\beta|^2 = 1$. The standard definition of degree of polarization yields $DOP = |2|\alpha|^2 -1|$, which is shown to coincide with the proposed method.}
    \label{fig:two}
\end{figure}

From the point of view of the polarization state alone, the state depicted in Fig.~\ref{fig:two} corresponds to partially polarized light. The polarization state of $\ket{\psi}$ is given by the reduced density operator obtained by tracing over the Hilbert space $\mathcal{H}_\text{o}$: 
\begin{equation}
\rho = \text{tr}_{\mathcal{H}_o}\ket{\psi}\bra{\psi} = \begin{pmatrix}
|\alpha|^2 & 0 \\ 
0 & 1-|\alpha|^2 \\
\end{pmatrix}
\label{eq:new2}
\end{equation}
As expected, the off-diagonal components of $\rho$ are zero, such that $s_2 = s_3 = 0$, and the degree of polarization is simply given by 
\begin{equation}
DOP = |s_1| = \text{tr}(\rho\sigma_1) = |2|\alpha|^2 -1|
\label{eq:new3}
\end{equation}
Now, consider the two-photon interference measurement of the state depicted in Fig.~\ref{fig:two}. It is straightforward to show that this is a particular case of Eq.~\ref{eq:one1} with $\xi = t_0$ and $\xi^\perp = t_1$, such that Eq.~\ref{eq:four4} now yields:
\begin{equation}
\mathcal{F}(U) = \left||\alpha|^2\braket{H|U|H} + |\beta|^2\braket{V|U|V}\right|^2
\label{eq:eff}
\end{equation}
which is clearly zero for $U\in\{\sigma_2,\sigma_3\}$ and equal to $|\alpha|^2-|\beta|^2 = |2|\alpha|^2-1|$ for $U = \sigma_1$, which coincides with the standard tomography result given by Eq.~\ref{eq:new3}. For a detailed calculation we refer the reader to Appendix \ref{app:b}.

As previously mentioned, when $U = I$, we have $\mathcal{F}(I) = 1$, which translates to a zero coincidence probability in the HOM setting, according to Eq.~\ref{eq:one}. This result, in fact, occurs independently of the way the polarization state is coupled to any inner degree of freedom of the photon. This happens because the photons remain in a pure state, and whenever two identical photons described by a pure quantum state impinge on a BS,  the probability amplitudes of one photon being transmitted and the other reflected (and vice-versa) are out of phase, such that the photon bunching effect occurs. This is a remarkable advantage of the proposed protocol: by measuring the coincidence probability $P_{coinc}(I)$, one can verify whether the incoming state is indeed a pure state. 

\subsection{\label{sec:ext_ent}``External" entanglement and incoherent mixtures of pure states}

If the coincidence probability $P_{coinc}(I) > 0$, the input states are not pure states; by "pure" state, we mean the density matrix that encompasses all degrees of freedom of the single photon can be represented by a rank 1 projector. If the coincidence rate is greater than zero, then two possibilities arise: first, the case where random time-varying unitary operations produce resulting mixed states, i.e, incoherent mixtures of pure states; second, and assuming that the two photons at the BS inputs have been collected from the SPS in a controllable way -- such that we can rule out such fast unitary operations --, the case where the  polarization state and an external degree of freedom are entangled. Examples of the latter are two photons that are each half of a polarization-entangled pair; or photons produced through the decay of an atomic level that retains information about the polarization state of the photon, as in certain Raman-based atom-photon entanglement protocols~\cite{jing2019entanglement}. 
In both cases described above, the input states in Fig.~\ref{fig:one} are now represented by a density operator $\rho$. Irrespective of the actual nature of $\rho$, we can always write:
\begin{equation}
\rho = \lambda\ket{\phi}\bra{\phi} + (1-\lambda)\ket{\phi^\perp}\bra{\phi^\perp}
\label{eq:ten}
\end{equation}
where the parameter $\lambda$ is directly related to the DOP = $|2\lambda-1|$. Now we can apply the following reasoning: whenever both inputs are identical, which occurs with probability $\lambda^2+(1-\lambda)^2$, the fidelity is one; however, when both inputs are different, which happens with probability $2\lambda(1-\lambda)$, the fidelity is zero. Therefore, the average fidelity is given by $\mathcal{F}(I) = \lambda^2+(1-\lambda)^2$, which corresponds to a coincidence probability given by:
\begin{equation}
P_{coinc}(I) = \lambda(1-\lambda).
\label{eq:eleven}
\end{equation}

For the Stokes parameters $\{s_j\}$, the calculation is somewhat different from the case of pure states, given by Eq.~\ref{eq:four4}. Indeed, Eq.~\ref{eq:four4} considers that all degrees of freedom are available when performing the inner product in Eq.~\ref{eq:two2}. Here, the fidelity is taken over two mixed states, i.e., between $\rho$ and $U\rho U^\dagger$. A simple calculation shows that:
\begin{align}
\mathcal{F}(\sigma_j) &= s_j^2 [\lambda^2+(1-\lambda)^2] + (1-s_j^2)2\lambda(1-\lambda)\nonumber\\
&= s_j^2(2\lambda-1)^2 + 2\lambda(1-\lambda)\label{eq:new_eq}.
\end{align}
Note that, whenever $\lambda = 0$ or $\lambda = 1$, Eq.~\ref{eq:new_eq} reduces to Eq.~\ref{eq:four4}, as expected. The probabilities of coincidence for a generalized state, replacing Eq. \ref{eq:new_eq} on Eq. \ref{eq:one}, now read:
\begin{align}
P_{coinc}(\sigma_j) &= \frac{1- s_j^2(2\lambda-1)^2}{2} - \lambda(1-\lambda)\nonumber\\
&= \frac{1-\langle s_j \rangle ^2}{2} - P_{coinc}(I),\label{eq:PcoincSigma_Mixed}
\end{align}
where $\langle s_j \rangle = s_j|2\lambda-1|$, a generalization of Eq.~\ref{eq:four}.

It is interesting to identify that $\lambda$ now has a two-fold effect on $P_{coinc}(\sigma_j)$: it acts on the first right hand side term of Eq.~\ref{eq:PcoincSigma_Mixed} similarly to $p$ in Eq.~\ref{eq:four4}, but the existence of external entanglement decreases all coincidence probabilities by $P_{coinc}(I)$, also governed by $\lambda$. Whenever $\lambda = \tfrac{1}{2}$, the probabilities are all equal to $\tfrac{1}{4}$, as one should expect: two completely mixed states impinging on the beamsplitter correspond to a 50-50 mixture of two identical states (zero coincidence probability) and two orthogonal states (50\% coincidence probability). Again, external entanglement is indistinguishable from incoherent mixtures of pure states; however, as long as random time-varying unitary operations are known to be not present, this protocol allows one to distinguish between ``inner" and ``outer" entanglement scenarios by computing $P_{coinc}(I)$ and verifying whether it is equal to zero (``inner" entanglement) or grater than zero (``outer" entanglement). Conversely, when entanglement with external degrees of freedom can be neglected or prevented, the protocol is able to distinguish between internal entanglement and incoherent mixtures. In both cases, these distinctions are unattainable in standard quantum state tomography.

Finally, it should be mentioned that both kinds of entanglement can be quantified, e.g., by calculating the von Neumann entropy of the reduced density matrix, $S(\rho)$, as usual. 

\subsection{Advantages over Standard QST}

The main advantage of the method stems from the combination of the calculated DOP from Eq. 8 and the coincidence probability $P_{coinc}(I)$. If $\rho$ is the density operator representing all (internal) degrees of freedom of the photon, then by definition  $P_{coinc}(I) = \text{tr}(\rho^2)$. This information, together with the DOP value, answers the following two questions at the same time: (i) ``Is there entanglement between the polarization state and another internal degree of freedom?'' and (ii) ``Is there entanglement between a internal degree of freedom of the photon and an external environment or, alternatively, an incoherent mixture prepared using classical probability sampling?''

Standard QST cannot directly provide these answers, unless all degrees of freedom are separately discriminated. 
Let us consider a simpler case where there are only two degrees of freedom of interest, as in subsection \ref{sec:int_ent}. 
Standard QST would need to reconstruct the 4x4 polarization-time-bin density matrix, which requires measurements of average values of all combinations of Pauli operators $\sigma_i\otimes\sigma_j$, which adds up to 16 measurement rounds; moreover, the detectors would have to be fast enough (and have low enough jitter) to be able to distinguish arrival times $t_0$ from $t_1$. 
The current method, however, needs only four measurement rounds and there is no constraint on the frequency response or the timing jitter of the detectors. 
This can be compared to an efficient QST, as the required number of resources does not scale exponentially with the system dimension~\cite{cramer2010}.

\subsection{Robustness to noise}

Up to this point we have considered perfect detectors and perfect single-photon sources. Therefore, a natural question arises: how robust to noise is the method? 
There are two main sources of noise that could influence the detection probabilities and potentially cause alterations in the measurement results. The first one are accidental counts in the detectors, which stem from the dark count probability density per second $p_{dc}$, the afterpulse probability $p_{a}$ and the background noise probability $p_{bkg}$. Assuming that $p_{bkg}$ can be completely eliminated by proper filtering and that $p_a$ can be made negligible by using a sufficiently long dead time, the dark counts will be the dominating source of accidental counts. In fact, accidental counts alone have no meaning: the figure of merit we are looking for is the signal-to-noise ratio, or in this case, the coincidence-to-accidental count ratio $R_{CA}$. This ratio also depends on other detection parameters, such as the detection window duration $\Delta T$ and the detection efficiency $\eta$.

The second main source of noise is the photon number probability distribution of the single photon source. Ideal sources produce Fock states with exactly one photon per pulse, but in practice there are small (usually negligible) probabilities of generating two or more photons or - much more frequently - zero photons, which corresponds to a vacuum state. The probability $p_{ph}$ that a pulse emitted by the source contains a photon defines the figure of merit that is relevant to this discussion.

It is straightforward to show that, whenever two (distinguishable) photons impinge on the BS, the coincidence-to-accidental count ratio is given by
\begin{equation}
    R_{CA} = \left(\frac{\eta p_{ph}}{p_{dc}\Delta T}\right)^2
    \label{eq:RCA}
\end{equation}

Let us now plug into Eq. \ref{eq:RCA} worst-case scenario numbers for existing single-photon detectors. Considering current off-the-shelf components, the single photon detection technology that yields the smaller quantum efficiency and highest dark count rate corresponds to InGaAs APDs for detection of telecommunication wavelength photons \cite{Gisin2002}. For this kind of detector, the dark count probability is at most $p_{dc} \approx 10^{-5}/ns$, and the quantum efficiency at least $\eta = 0.1$; see, e.g., idQuantique's detectors. The detection gate duration $\Delta T$ must correspond to at least the coherence time of the photons, which are at most in the order of 1ns for quantum-dot sources \cite{Nawrath2023,Arakawa2020}. Even taking an extreme value of $\Delta T = 10ns$, we obtain $R_{CA} \approx 10^6$ for a perfect single-photon source. As one can immediately conclude, even if the source deviates by a factor 10 from an ideal one (i.e., $p_{ph} = 0.1$), the coincidence count rate due to legitimate photons would still be four orders of magnitude above the accidental counts in the worst case scenario, i.e., using noisy and low-efficiency detectors. For this reason, we neglect noise counts in the next section while performing the simulations.\\

\section{Simulations}
In this section we provide simulations for the proposed protocol in a simulated quantum computer using the Qiskit package~\cite{Qiskit}. 
The basic idea is to use the equivalence between the HOM effect and the so-called Swap Test~\cite{quantumfp}, which has already been studied in~\cite{hom_swaptest,swaptest-puc} for single qubits, i.e., for a single degree of freedom, but also for multimode input states~\cite{Foulds_2021}, i.e., a bi-partite entangled state. Indeed, a HOM interference effect between two pure polarization states is equivalent to asking ``are the two polarization states identical?'', which, in its turn, is the question answered by the projection over the singlet state $\ket{\psi^-}$.

Before performing the experiments, 10000 states over the Bloch Sphere were randomly selected to create an uniform input data for all assessments.

\subsection{\label{sec4A}Polarized single photon states}

\begin{figure}[t]
    \centering
    \includegraphics[width=\linewidth]{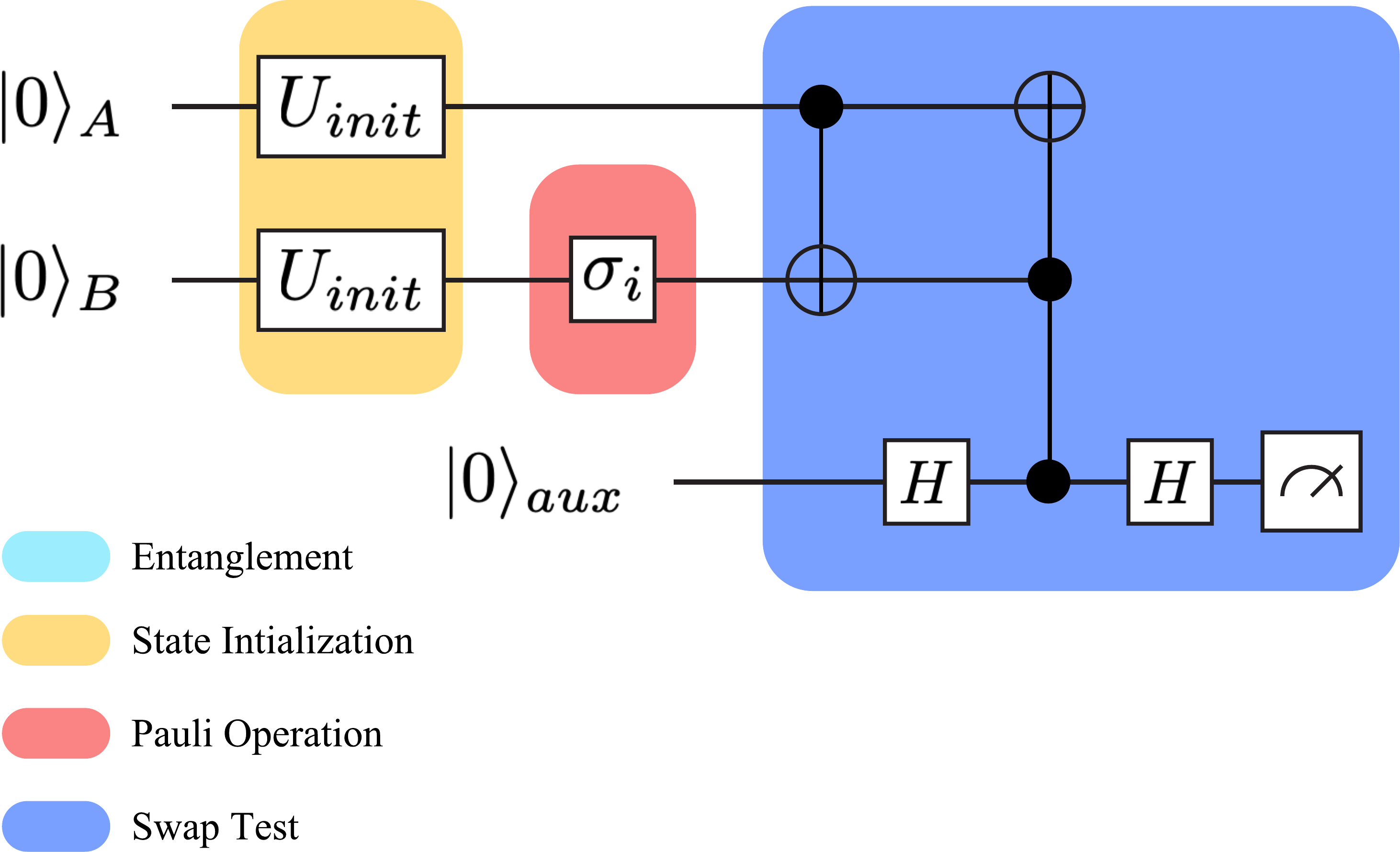}
    \caption{Swap Test quantum circuit using a Toffoli gate. By measuring the ancilla qubit in mode \textit{aux}, it is possible to determine whether states in modes \textit{A} and \textit{B} are identical, which mimics the statistics of a HOM interferometer. Figure adapted from Ref.~\cite{swaptest-puc}}.
    \label{fig:swaptest1}
\end{figure}

In the case of Fock states ($n = 1$) in pure polarization states, the HOM interferometer, presented in Fig.~\ref{fig:one}, is equivalent to a standard Swap Test, as shown in Fig.~\ref{fig:swaptest1}. In our case, qubit \textit{A} corresponds to state $\ket{\psi}$ and qubit \textit{B} to state $U\ket{\psi}$. Whenever qubits \textit{A} and \textit{B} are identical (which happens when $U = I$), measurement of the ancilla qubit \textit{C} always result in a projection onto state $\ket{0}$: 
\begin{equation}
Prob(``0") = \frac{1}{2}(1+|\braket{\psi | U | \psi}|^2) = 100\%
\label{eq:twelve}
\end{equation}
which is equivalent to a coincidence measurement $P_{coinc}(I)$ in a HOM scenario.

As presented in Fig.~\ref{fig:swaptest1}, in the simulation we prepare the aforementioned arbitrary states, with previously known Stokes parameters, and use the method to estimate $P_{coinc}^{exp}(\sigma_j)$. The random states are obtained by a $U_{init}$ operator, comprised of a $R_Y(\theta)$ and $R_Z(\phi)$ gates, where $\theta$ and $\phi$ are calculated according to the random state's Stokes vector:

\begin{align}
    \overrightarrow{s} &= (s_1,s_2,s_3) \nonumber \\
   &=  DOP(\cos \theta , \cos \phi \sin \theta, \sin \phi \sin \theta )
   \label{eq:stokesvec}
\end{align}

Note that, for simplicity, we keep the original definition of the Stokes parameters such that $s_1 = \braket{\psi | Z | \psi}, s_2 = \braket{\psi | X | \psi}$ and $s_3 = \braket{\psi | Y | \psi}$.
Additionally, in this case, Eq.~\ref{eq:stokesvec} will have $DOP = 1$ for all randomly selected states - in the next section we will deal with partially polarized light.

We now define the Stokes vector error estimation parameter $\varepsilon$ as the square sum of the differences between the theoretical and experimental values of the coincidence probabilities for all Pauli operators, as in  Eq.~\ref{eq:error_formula}: 
\begin{align}
\varepsilon &\equiv \sum_{j=1}^3\left|P_{coinc}(I)+P_{coinc}(\sigma_j) -P_{coinc}^{exp}(I)-P_{coinc}^{exp}(\sigma_j)\right|^2\nonumber\\
&= \sum_{j=1}^3\left|\frac{1-\langle s_j\rangle ^2}{2}-P_{coinc}^{exp}(I)-P_{coinc}^{exp}(\sigma_j)\right|^2
\label{eq:error_formula}
\end{align}
where Eq. \ref{eq:PcoincSigma_Mixed} has been employed and \textit{exp} stands for "experimental". It is worth mentioning that for this experiment and the following one, which are not comprised of external entanglement, we expect to find near-zero values for $P_{coinc}^{exp}(I)$.

\begin{figure}[ht]
    \centering
    \includegraphics[width = \linewidth]{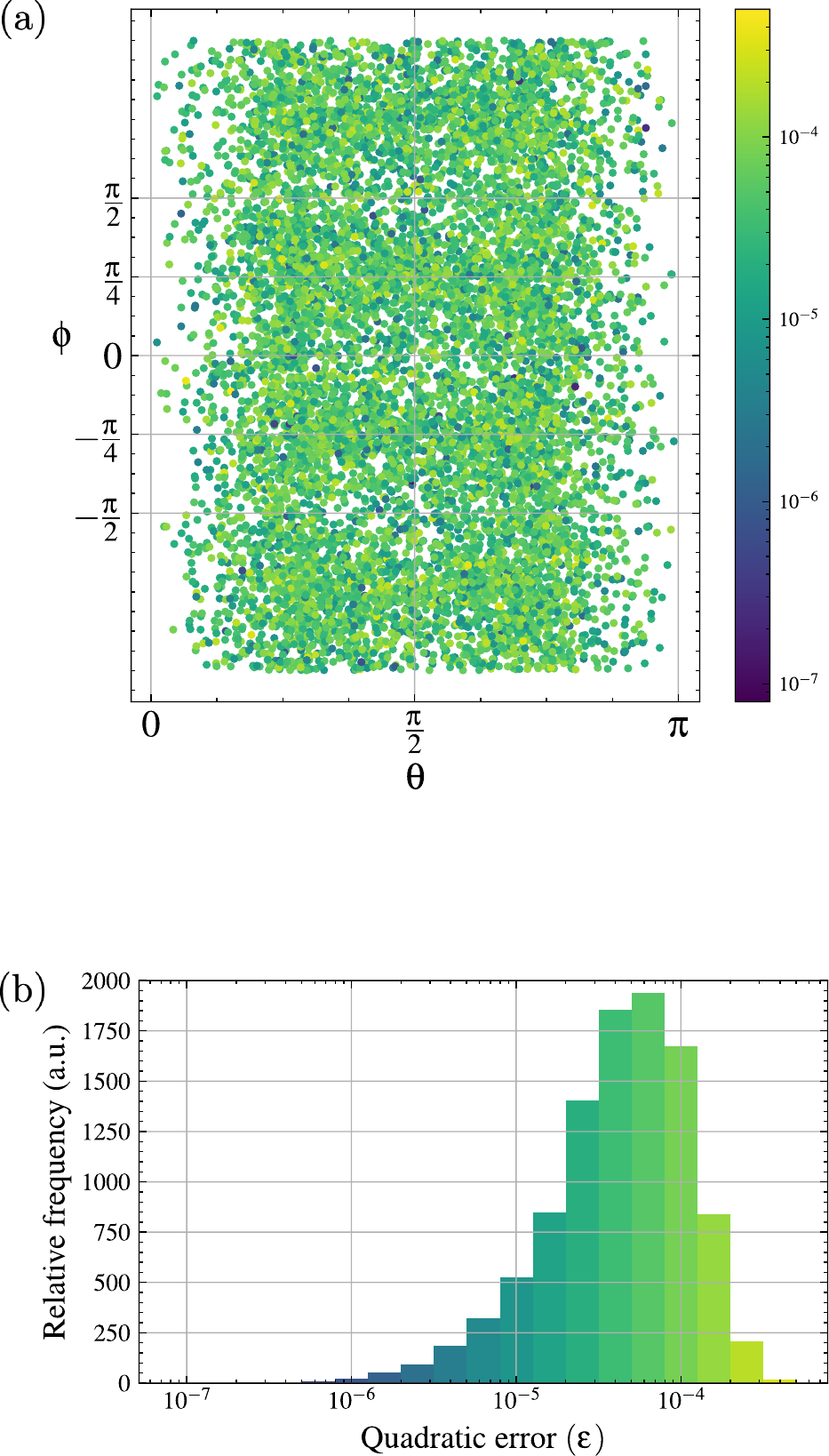}
    \caption{(a) Heatmap scatter plot of the proposed method's error $\varepsilon$, as defined in Eq.~\ref{eq:error_formula}, versus the polar coordinates of the input state. For the pure state input considered in this Section, all states lie on the shell of the Poincaré Sphere, which enables this representation.(b) Histogram of the error magnitude relative frequency, indicating that the errors lie mostly in the range between $10^{-5}$ and $10^{-4}$.}
    \label{fig:exp_1_error}
\end{figure}

The estimated error according to Eq.~\ref{eq:error_formula} for each arbitrary state is depicted in Fig.~\ref{fig:exp_1_error}.
It is perceivable that our method is reliable, regardless of the quantum state at hand, as there is no correlation between error and position in the Bloch Sphere. The wide range in orders of magnitude is likely due to statistical fluctuations on the estimation of the probabilities and rounding errors.

\subsection{Single photon states with internal entanglement}
\begin{figure}[t!]
    \centering
    \includegraphics[width=\linewidth]{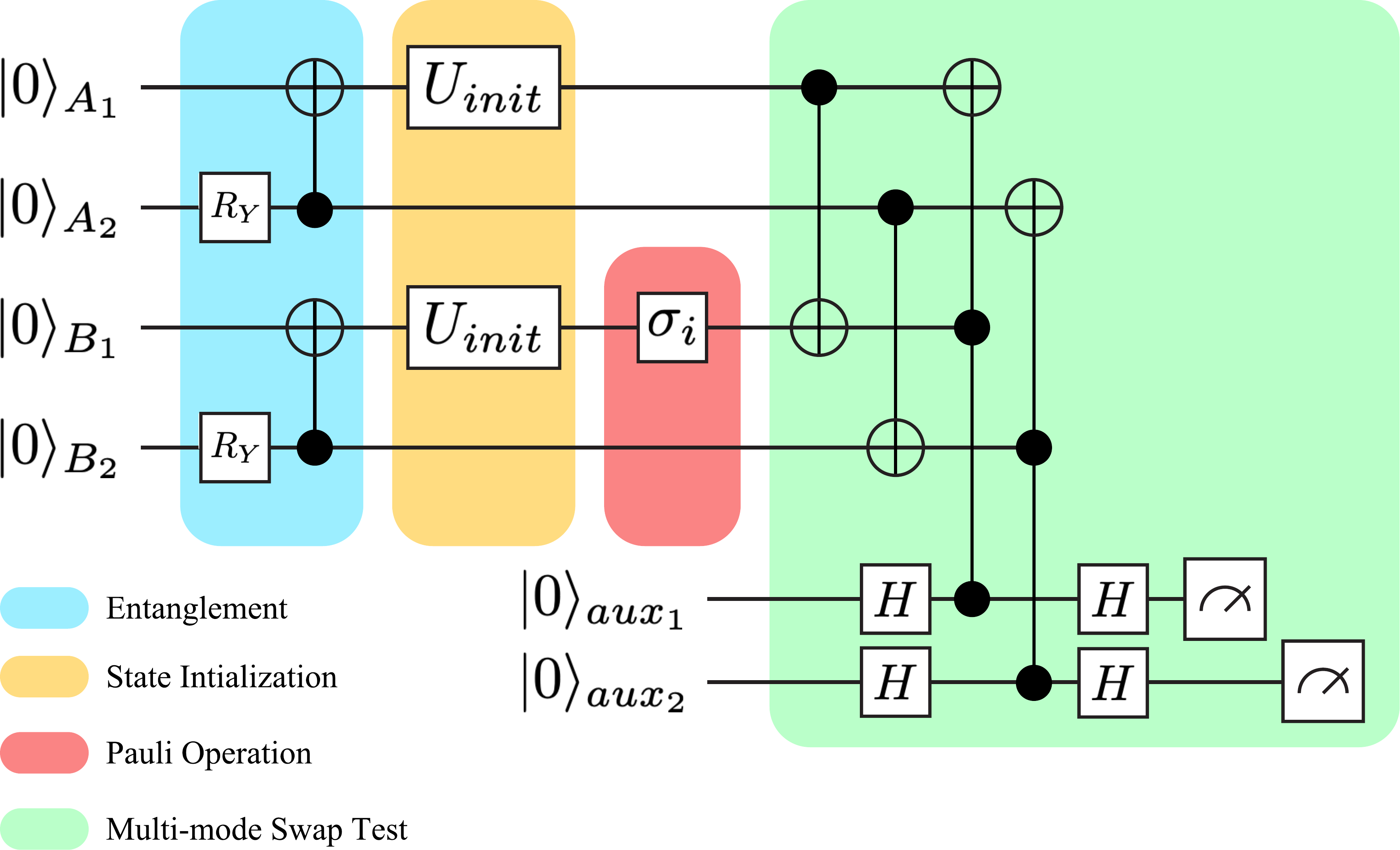}
    \caption{Swap Test quantum circuit for two-mode inputs. Each degree of freedom is simulated by a qubit. This circuit uses two ancilla qubits, and the probability of obtaining a result of ``01" or ``10" is equivalent to a coincidence measurement in a HOM interferometer. Adapted from Ref.~\cite{Foulds_2021}.}
    \label{fig:swaptest2}
\end{figure}

Now we consider the case of two internal degrees of freedom, such as in the example of Fig.~\ref{fig:two}. In this case, we employ the circuit shown in Fig.~\ref{fig:swaptest2}. Without loss of generality, we can consider the cases where the states are given by, up to a global unitary operation:
\begin{eqnarray}
\ket{\psi}_A = && \alpha_0\ket{00} + \alpha_1\ket{11} \nonumber \\
\ket{\phi}_B = && \beta_0\ket{00} + \beta_1\ket{11}
\label{eq:thirteen}
\end{eqnarray}
where $|\alpha_0|^2+|\alpha_1|^2 = 1$ and $|\beta_0|^2+|\beta_1|^2 = 1$. It is possible to show, using the results from Foulds \textit{et al}~\cite{Foulds_2021}, that the probability of obtaining a result ``01" or ``10" on the measurement of the ancillas is given by
\begin{equation}
\text{Prob}(``01",``10") = \frac{1}{2}|\alpha_0\beta_1 - \alpha_1\beta_0|^2
\label{eq:fourteen}
\end{equation}

Note that, in agreement with the HOM interferometer, in Fig.~\ref{fig:one}, whenever qubits \textit{A} and \textit{B} are identical ($\alpha_i = \beta_i$), the probability in Eq.~\ref{eq:fourteen} is zero, which corresponds to the zero coincidence probability. Moreover, if \textit{A} and \textit{B} are mutually orthogonal ($\alpha_0 = \beta_1^*, \alpha_1 = -\beta_0^*$) then the probability becomes 1/2, i.e., the photons are completely distinguishable.

\begin{figure}[ht]
    \centering
    \includegraphics[width = \linewidth]{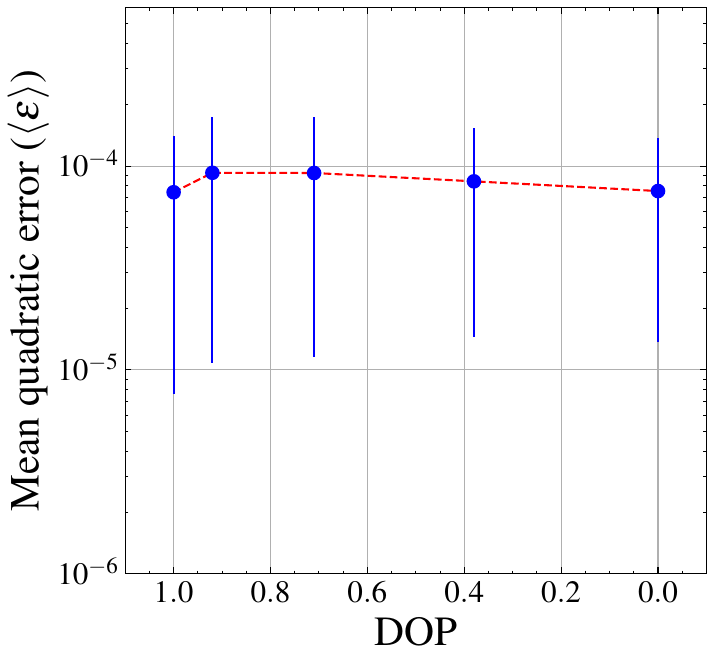}
    \caption{Mean quadratic polarimetry error, simulated in a quantum computer emulator using the schematic of Fig.~\ref{fig:swaptest2}, as a function of the DOP of the input states. The mean was calculated over the 10000 random states.}
    \label{fig:exp_2_error}
\end{figure}

The formula for the error estimation parameter $\varepsilon$ from Eq.~\ref{eq:error_formula} was also applied to this experiment, with a small adaptation. 
In the first experiment, only pure states were considered, while this case concerns mixed states (from the point of view of the polarization degree of freedom). 
In order to generate such states, we decreased the norm of every random pure state, i.e. a Stokes vector with unit norm.
This was performed by multiplying the vector by the DOP. The resulting values of the Stokes parameters $s_j$ were then obtained as follows:
\begin{align}
    s_j = & s^{pure}_j (DOP_{theory}), \text{where} \\
    DOP_{theory} = & \sqrt{1-4 \text{det}(\rho)}
\end{align}
where $s_j^{pure}$ are the coordinates of the initial random Bloch vector and $\rho$ is the reduced density matrix tracing over all other degrees of freedom other than polarization. The results are shown in Fig.~\ref{fig:exp_2_error}, which portrays the resulting mean quadratic error for different DOP values. Each point in this graph corresponds to an average taken over 10000 points with 30000 measurement rounds~($\# shots * 3$) for each measurement round of each point. Please note that the error bars in Fig. \ref{fig:exp_2_error} (and also Fig. \ref{fig:exp_3_error}) are symmetric, but distorted by the logarithmic scale.

\subsection{Single photon states with external entanglement}

\begin{figure}[b!]
    \centering
    \includegraphics[width=\linewidth]{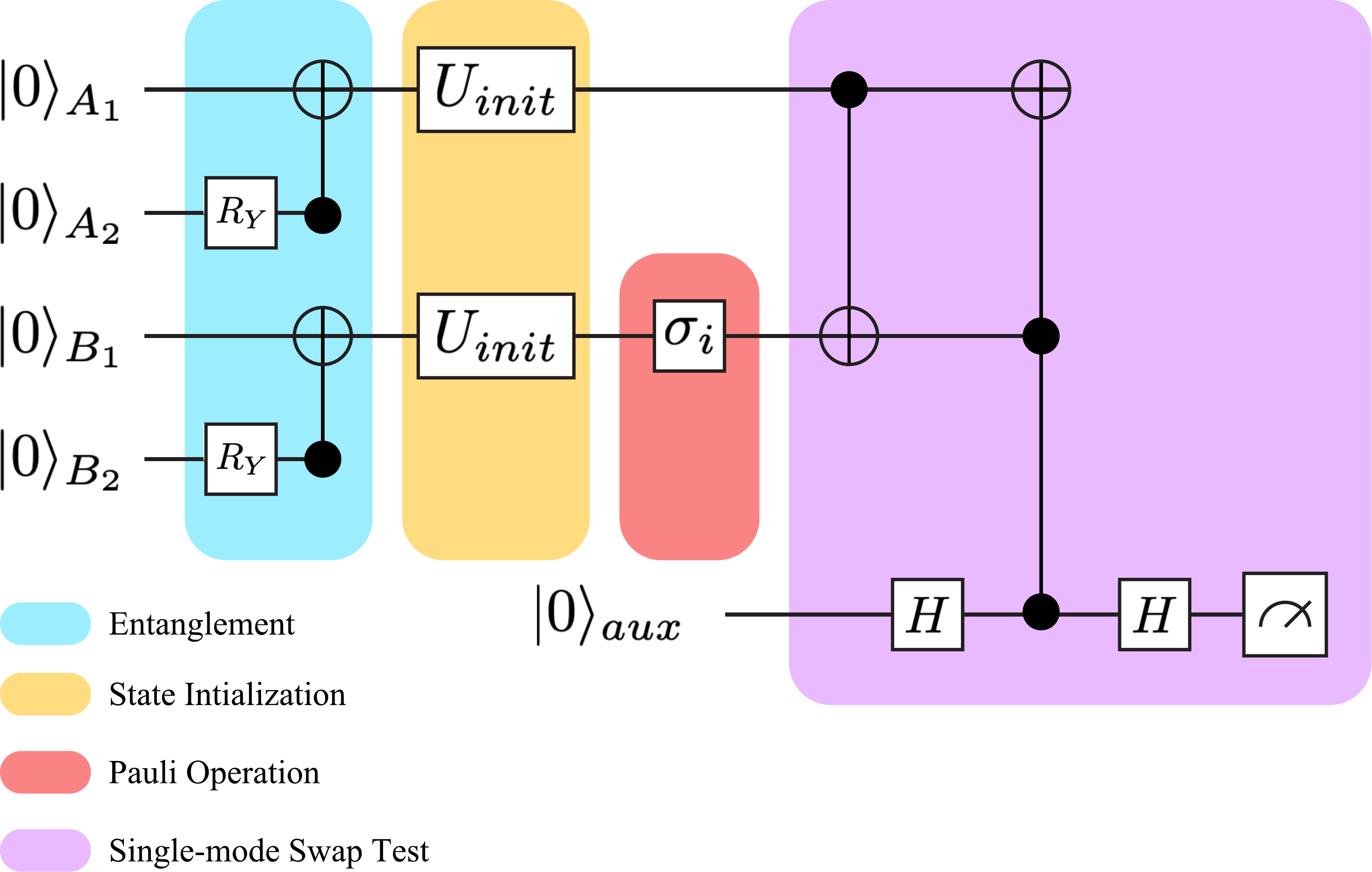}
    \caption{Swap Test quantum circuit for single-mode inputs where each input is entangled with an external and inaccessible degree of freedom (the ``environment"). Adapted from Ref.~\cite{swaptest-puc}.}
    \label{fig:swaptest3}
\end{figure}

In this section we deal with the situations where the incoming single photons are mixed states, in the sense that the polarization states are entangled with an external degree of freedom - the ``environment". For simplicity, we assume a single-dimensional environment for each input qubit. The preparation procedure of the states is very similar to the internal entanglement case, but the measurement is identical to the first case (pure polarization states). The setup can be found in Fig.~\ref{fig:swaptest3}.

The quantum circuit of Fig.~\ref{fig:swaptest3} is essentially the same one from Fig.~\ref{fig:swaptest1}, with the addition of two qubits representing the environment. The probability of measuring ``0" in the ancilla qubit $C$ is equivalent to projecting the two-qubit input state over the singlet state $\ket{\psi^-}$, such that
\begin{equation}
\text{Prob}(``0") = \braket{\psi^- | \rho_A \otimes \rho_B | \psi^- }
\label{eq:twenty-two}
\end{equation}
where $\rho_{A,B}$ are the reduced density matrices obtained by tracing over the environment. Whenever the two input qubits are the same, which corresponds to the situation where $U = I$, a simple calculation shows that, whenever $\rho_A = \rho_B = \rho$, then
\begin{equation}
\braket{\psi^- | \rho \otimes \rho | \psi^- } = \text{det}(\rho)
\label{eq:twenty-three}
\end{equation}
which mimics the result previously given by Eq.~\ref{eq:eleven}. 

\begin{figure}[t!]
    \centering
    \includegraphics[width = \linewidth]{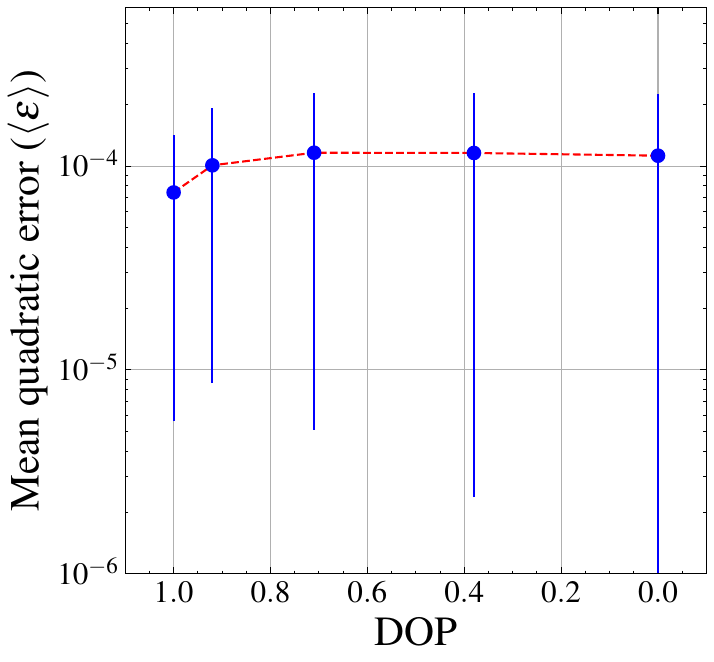}
    \caption{Mean quadratic polarimetry error, simulated in a quantum computer emulator using the schematic of Fig.~\ref{fig:swaptest3}, as a function of the DOP of the input states. The mean was calculated over the 10000 randomly selected states.}
    \label{fig:exp_3_error}
\end{figure}

As demonstrated in Figures \ref{fig:exp_2_error} and \ref{fig:exp_3_error}, the proposed method is able to correctly characterize the input state independently of the degree of entanglement of the photons, be it internal or external.

\subsection{Benchmarking with standard QST}

\begin{figure}[b!]
    \centering
    \includegraphics[width = \linewidth]{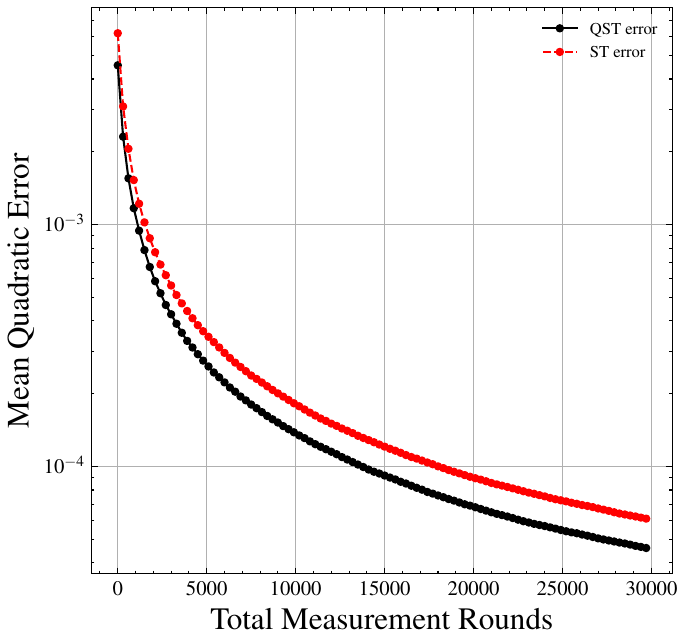}
    \caption{Mean quadratic error for the 10000 randomly selected states, according to the total number of measurement rounds~($\#shots * 3)$ performed for each point.
    QST: Standard Quantum State Tomography.
    ST: Swap Test (proposed method)}
    \label{fig:benchmark_1}
\end{figure}

The results shown in Figs.~\ref{fig:exp_1_error}, \ref{fig:exp_2_error}~and~\ref{fig:exp_3_error} demonstrate that the proposed method performs similarly independently on the input state; in other words, the robustness of the method is validated against itself. In order to extend this analysis, we introduce the standard quantum state tomography (QST) as a reference for the performance. Without loss of generality, we perform an extensive test using pure state inputs, i.e., those discussed in Section \ref{sec4A}, on the standard QST and the method depicted as a block diagram in Fig.~\ref{fig:swaptest1}. The approach to the evaluation of the performance is the same: the input states are varied to map different points on the shell of the Poincaré Sphere; but, now, the number of measurement rounds (or realization of the same input state) is also varied; for each number of measurement rounds, the error averaged over all input states is computed following Eq.~\ref{eq:error_formula}. The results, presented in Fig.~\ref{fig:benchmark_1}, demonstrate that the error achieved by the proposed method and the standard QST method is comparable, which validates the two-photon quantum state tomography of photonic qubits. Moreover, the standard QST implementation on Qiskit introduces a step of regularization to ensure physically meaningful tomographic results drawn from a finite set of measurements. Due to this extra data processing step, the QST is likely to achieve smaller errors, which is clearly reproduced in the results depicted in Figure \ref{fig:benchmark_1}.

It is worth noting that the results in Fig.~\ref{fig:benchmark_1} would not be attainable in practice. For the standard QST, errors smaller than $10^{-4}$, would only be achievable in case a polarizing beamsplitter (PBS) with an extinction ratio larger than 40 dB was available, which is very far from actual commercial values of $\sim$ 20-27 dB~\cite{Calliari:19}. On the other hand, the proposed method of two-photon polarimetry does not resort to polarization-selective components, but the results correspond to a perfectly symmetrical beamsplitter (BS) being employed. A deviation of about 0.5 dB is expected between the BS outputs; however, this deviation can be compensated by adding the corresponding loss in its less-affected output, meaning that the mean quadratic error $\varepsilon$ is only limited by calibration errors of the BS asymmetry, which can be made as low as one requires.

\section{Generalization of the method}

Throughout the paper we have considered that the polarization state is to be determined; however, the method can be equally applied to characterize any two-level encoding of quantum information in single photons. The unitary operators represented by $U$ in Fig.~\ref{fig:one} can straightforwardly be replaced by Pauli operators that act on any two-dimensional Hilbert space, henceforth performing a characterization of any other degree of freedom, such as time-bin, spatial mode~\cite{PhysRevA.88.012302}, orbital angular momentum~\cite{PhysRevA.83.060301} or frequency encoding~\cite{frequency_bins}. Of course, the meaning of "DOP" would have to be adjusted accordingly - generic terms such as "mutual coherence" may be employed for these other cases, always corresponding to the norm of the Bloch/Stokes vector in a two-dimensional Hilbert space.

Moreover, the ``polarization-dependent loss" introduced in Appendix \ref{app:a}
would have to be changed accordingly. For example, for time-bin qubits, a time-dependent loss element would be needed. Again, the actual value of the loss does not have any impact over the results.

Finally, a remarkable advantage of our method is that it allows the evaluation of degrees of freedom that would otherwise be inaccessible by direct measurement. For example, if the Pauli operators $\sigma_j$ act on the time-bin Hilbert space, the purity of the time-bin qubit and all its Bloch sphere components can be determined even if the time delay between the two bins are shorter than the timing resolution of the detectors, whereas this degree of freedom would be traced out in any direct measurement of individual single photons.

\section{Conclusion}

We have presented a method for performing full quantum state tomography in qubits encoded in single photons, such as the polarization degree of freedom, that relies on the Hong-Ou-Mandel effect. By interfering the original quantum state in a beamsplitter with the image of the input state by each of the three Pauli spin operators, and by measuring the coincidence count rates between single photon detectors placed at each beamsplitter output, the absolute values of the Bloch/Stokes parameters can be obtained. By introducing three additional measurements in rotated versions of the original state, the information on the signs of the Bloch/Stokes parameters can be found. Moreover, by simply interfering the unknown photon with a copy of itself, the protocol is able to tell whether the polarization state is entangled with another internal degree of freedom, such as time-bin, or entangled with an external degree of freedom, as long as random time-varying unitary operations or classical probability samplings can be ruled out. This is an advantage over standard QST schemes that would not only need to have access and the ability to resolve these degrees of freedom but also introduce additional measurements that scale exponentially with the number of degrees of freedom under consideration. Simulations using quantum circuits in a Swap-Test configuration show that the errors introduced by the method are negligible, on the same order of magnitude as standard QST, provided that a sufficiently large number of copies are measured.

For polarization qubits, we have shown that this method does not rely on polarization-selective components, such as polarizers or polarizing beamsplitters, which means that the measurements are not limited by the extinction ratio of any such components. Therefore, we believe that the presented method can be useful in measuring the degree of polarization of very pure states. 

\begin{acknowledgments}
G. P. Temporão and T. B. Guerreiro acknowledge financial support from FAPERJ, FAPESP and CNPq.

\end{acknowledgments}

\bibliography{apssamp.bib}

\providecommand{\noopsort}[1]{}\providecommand{\singleletter}[1]{#1}%
\begin{thebibliography}{34}%
\makeatletter
\providecommand \@ifxundefined [1]{%
 \@ifx{#1\undefined}
}%
\providecommand \@ifnum [1]{%
 \ifnum #1\expandafter \@firstoftwo
 \else \expandafter \@secondoftwo
 \fi
}%
\providecommand \@ifx [1]{%
 \ifx #1\expandafter \@firstoftwo
 \else \expandafter \@secondoftwo
 \fi
}%
\providecommand \natexlab [1]{#1}%
\providecommand \enquote  [1]{``#1''}%
\providecommand \bibnamefont  [1]{#1}%
\providecommand \bibfnamefont [1]{#1}%
\providecommand \citenamefont [1]{#1}%
\providecommand \href@noop [0]{\@secondoftwo}%
\providecommand \href [0]{\begingroup \@sanitize@url \@href}%
\providecommand \@href[1]{\@@startlink{#1}\@@href}%
\providecommand \@@href[1]{\endgroup#1\@@endlink}%
\providecommand \@sanitize@url [0]{\catcode `\\12\catcode `\$12\catcode
  `\&12\catcode `\#12\catcode `\^12\catcode `\_12\catcode `\%12\relax}%
\providecommand \@@startlink[1]{}%
\providecommand \@@endlink[0]{}%
\providecommand \url  [0]{\begingroup\@sanitize@url \@url }%
\providecommand \@url [1]{\endgroup\@href {#1}{\urlprefix }}%
\providecommand \urlprefix  [0]{URL }%
\providecommand \Eprint [0]{\href }%
\providecommand \doibase [0]{https://doi.org/}%
\providecommand \selectlanguage [0]{\@gobble}%
\providecommand \bibinfo  [0]{\@secondoftwo}%
\providecommand \bibfield  [0]{\@secondoftwo}%
\providecommand \translation [1]{[#1]}%
\providecommand \BibitemOpen [0]{}%
\providecommand \bibitemStop [0]{}%
\providecommand \bibitemNoStop [0]{.\EOS\space}%
\providecommand \EOS [0]{\spacefactor3000\relax}%
\providecommand \BibitemShut  [1]{\csname bibitem#1\endcsname}%
\let\auto@bib@innerbib\@empty
\bibitem [{\citenamefont {Gisin}\ and\ \citenamefont {Thew}(2007)}]{Rob2007}%
  \BibitemOpen
  \bibfield  {author} {\bibinfo {author} {\bibfnamefont {N.}~\bibnamefont
  {Gisin}}\ and\ \bibinfo {author} {\bibfnamefont {R.}~\bibnamefont {Thew}},\
  }\href {https://doi.org/10.1038/nphoton.2007.22} {\bibfield  {journal}
  {\bibinfo  {journal} {Nature Photonics}\ }\textbf {\bibinfo {volume} {1}},\
  \bibinfo {pages} {165} (\bibinfo {year} {2007})}\BibitemShut {NoStop}%
\bibitem [{\citenamefont {Couteau}\ \emph {et~al.}(2023)\citenamefont
  {Couteau}, \citenamefont {Barz}, \citenamefont {Durt}, \citenamefont
  {Gerrits}, \citenamefont {Huwer}, \citenamefont {Prevedel}, \citenamefont
  {Rarity}, \citenamefont {Shields},\ and\ \citenamefont
  {Weihs}}]{Couteau2023}%
  \BibitemOpen
  \bibfield  {author} {\bibinfo {author} {\bibfnamefont {C.}~\bibnamefont
  {Couteau}}, \bibinfo {author} {\bibfnamefont {S.}~\bibnamefont {Barz}},
  \bibinfo {author} {\bibfnamefont {T.}~\bibnamefont {Durt}}, \bibinfo {author}
  {\bibfnamefont {T.}~\bibnamefont {Gerrits}}, \bibinfo {author} {\bibfnamefont
  {J.}~\bibnamefont {Huwer}}, \bibinfo {author} {\bibfnamefont
  {R.}~\bibnamefont {Prevedel}}, \bibinfo {author} {\bibfnamefont
  {J.}~\bibnamefont {Rarity}}, \bibinfo {author} {\bibfnamefont
  {A.}~\bibnamefont {Shields}},\ and\ \bibinfo {author} {\bibfnamefont
  {G.}~\bibnamefont {Weihs}},\ }\href
  {https://doi.org/10.1038/s42254-023-00589-w} {\bibfield  {journal} {\bibinfo
  {journal} {Nature Reviews Physics}\ }\textbf {\bibinfo {volume} {5}},\
  \bibinfo {pages} {354} (\bibinfo {year} {2023})}\BibitemShut {NoStop}%
\bibitem [{\citenamefont {O'Brien}(2007)}]{OBrien07}%
  \BibitemOpen
  \bibfield  {author} {\bibinfo {author} {\bibfnamefont {J.~L.}\ \bibnamefont
  {O'Brien}},\ }\href@noop {} {\bibfield  {journal} {\bibinfo  {journal}
  {Science}\ }\textbf {\bibinfo {volume} {318}},\ \bibinfo {pages} {1567}
  (\bibinfo {year} {2007})}\BibitemShut {NoStop}%
\bibitem [{\citenamefont {Altepeter}\ \emph {et~al.}(2005)\citenamefont
  {Altepeter}, \citenamefont {Jeffrey},\ and\ \citenamefont
  {Kwiat}}]{Altepeter05}%
  \BibitemOpen
  \bibfield  {author} {\bibinfo {author} {\bibfnamefont {J.}~\bibnamefont
  {Altepeter}}, \bibinfo {author} {\bibfnamefont {E.}~\bibnamefont {Jeffrey}},\
  and\ \bibinfo {author} {\bibfnamefont {P.}~\bibnamefont {Kwiat}}\ }(\bibinfo
  {publisher} {Academic Press},\ \bibinfo {year} {2005})\ pp.\ \bibinfo {pages}
  {105--159}\BibitemShut {NoStop}%
\bibitem [{\citenamefont {Resch}\ \emph {et~al.}(2005)\citenamefont {Resch},
  \citenamefont {Walther},\ and\ \citenamefont {Zeilinger}}]{Resch05}%
  \BibitemOpen
  \bibfield  {author} {\bibinfo {author} {\bibfnamefont {K.~J.}\ \bibnamefont
  {Resch}}, \bibinfo {author} {\bibfnamefont {P.}~\bibnamefont {Walther}},\
  and\ \bibinfo {author} {\bibfnamefont {A.}~\bibnamefont {Zeilinger}},\ }\href
  {https://doi.org/10.1103/PhysRevLett.94.070402} {\bibfield  {journal}
  {\bibinfo  {journal} {Phys. Rev. Lett.}\ }\textbf {\bibinfo {volume} {94}},\
  \bibinfo {pages} {070402} (\bibinfo {year} {2005})}\BibitemShut {NoStop}%
\bibitem [{\citenamefont {Takesue}\ and\ \citenamefont
  {Noguchi}(2009)}]{Takesue09}%
  \BibitemOpen
  \bibfield  {author} {\bibinfo {author} {\bibfnamefont {H.}~\bibnamefont
  {Takesue}}\ and\ \bibinfo {author} {\bibfnamefont {Y.}~\bibnamefont
  {Noguchi}},\ }\href {https://doi.org/10.1364/OE.17.010976} {\bibfield
  {journal} {\bibinfo  {journal} {Opt. Express}\ }\textbf {\bibinfo {volume}
  {17}},\ \bibinfo {pages} {10976} (\bibinfo {year} {2009})}\BibitemShut
  {NoStop}%
\bibitem [{\citenamefont {Titchener}\ \emph {et~al.}(2018)\citenamefont
  {Titchener}, \citenamefont {Gr{\"a}fe}, \citenamefont {Heilmann},
  \citenamefont {Solntsev}, \citenamefont {Szameit},\ and\ \citenamefont
  {Sukhorukov}}]{Titchener18}%
  \BibitemOpen
  \bibfield  {author} {\bibinfo {author} {\bibfnamefont {J.~G.}\ \bibnamefont
  {Titchener}}, \bibinfo {author} {\bibfnamefont {M.}~\bibnamefont
  {Gr{\"a}fe}}, \bibinfo {author} {\bibfnamefont {R.}~\bibnamefont {Heilmann}},
  \bibinfo {author} {\bibfnamefont {A.~S.}\ \bibnamefont {Solntsev}}, \bibinfo
  {author} {\bibfnamefont {A.}~\bibnamefont {Szameit}},\ and\ \bibinfo {author}
  {\bibfnamefont {A.~A.}\ \bibnamefont {Sukhorukov}},\ }\href
  {https://doi.org/10.1038/s41534-018-0063-5} {\bibfield  {journal} {\bibinfo
  {journal} {npj Quantum Information}\ }\textbf {\bibinfo {volume} {4}},\
  \bibinfo {pages} {19} (\bibinfo {year} {2018})}\BibitemShut {NoStop}%
\bibitem [{\citenamefont {Peters}\ \emph {et~al.}(2003)\citenamefont {Peters},
  \citenamefont {Altepeter}, \citenamefont {Jeffrey}, \citenamefont
  {Branning},\ and\ \citenamefont {Kwiat}}]{Peters2003}%
  \BibitemOpen
  \bibfield  {author} {\bibinfo {author} {\bibfnamefont {N.}~\bibnamefont
  {Peters}}, \bibinfo {author} {\bibfnamefont {J.}~\bibnamefont {Altepeter}},
  \bibinfo {author} {\bibfnamefont {E.}~\bibnamefont {Jeffrey}}, \bibinfo
  {author} {\bibfnamefont {D.}~\bibnamefont {Branning}},\ and\ \bibinfo
  {author} {\bibfnamefont {P.}~\bibnamefont {Kwiat}},\ }\href@noop {}
  {\bibfield  {journal} {\bibinfo  {journal} {Quantum Information and
  Computation}\ }\textbf {\bibinfo {volume} {3}},\ \bibinfo {pages} {503}
  (\bibinfo {year} {2003})}\BibitemShut {NoStop}%
\bibitem [{\citenamefont {Hong}\ \emph {et~al.}(1987)\citenamefont {Hong},
  \citenamefont {Ou},\ and\ \citenamefont {Mandel}}]{HOM_original}%
  \BibitemOpen
  \bibfield  {author} {\bibinfo {author} {\bibfnamefont {C.~K.}\ \bibnamefont
  {Hong}}, \bibinfo {author} {\bibfnamefont {Z.~Y.}\ \bibnamefont {Ou}},\ and\
  \bibinfo {author} {\bibfnamefont {L.}~\bibnamefont {Mandel}},\ }\href@noop {}
  {\bibfield  {journal} {\bibinfo  {journal} {Phys. Rev. Lett.}\ }\textbf
  {\bibinfo {volume} {59}},\ \bibinfo {pages} {2044} (\bibinfo {year}
  {1987})}\BibitemShut {NoStop}%
\bibitem [{\citenamefont {Adamson}\ \emph {et~al.}(2007)\citenamefont
  {Adamson}, \citenamefont {Shalm},\ and\ \citenamefont
  {Steinberg}}]{steinberg}%
  \BibitemOpen
  \bibfield  {author} {\bibinfo {author} {\bibfnamefont {R.~B.~A.}\
  \bibnamefont {Adamson}}, \bibinfo {author} {\bibfnamefont {L.~K.}\
  \bibnamefont {Shalm}},\ and\ \bibinfo {author} {\bibfnamefont {A.~M.}\
  \bibnamefont {Steinberg}},\ }\href
  {https://doi.org/10.1103/PhysRevA.75.012104} {\bibfield  {journal} {\bibinfo
  {journal} {Phys. Rev. A}\ }\textbf {\bibinfo {volume} {75}},\ \bibinfo
  {pages} {012104} (\bibinfo {year} {2007})}\BibitemShut {NoStop}%
\bibitem [{\citenamefont {Ekert}\ \emph {et~al.}(2002)\citenamefont {Ekert},
  \citenamefont {Alves}, \citenamefont {Oi}, \citenamefont {Horodecki},
  \citenamefont {Horodecki},\ and\ \citenamefont {Kwek}}]{Ekert2002}%
  \BibitemOpen
  \bibfield  {author} {\bibinfo {author} {\bibfnamefont {A.~K.}\ \bibnamefont
  {Ekert}}, \bibinfo {author} {\bibfnamefont {C.~M.}\ \bibnamefont {Alves}},
  \bibinfo {author} {\bibfnamefont {D.~K.~L.}\ \bibnamefont {Oi}}, \bibinfo
  {author} {\bibfnamefont {M.}~\bibnamefont {Horodecki}}, \bibinfo {author}
  {\bibfnamefont {P.}~\bibnamefont {Horodecki}},\ and\ \bibinfo {author}
  {\bibfnamefont {L.~C.}\ \bibnamefont {Kwek}},\ }\href
  {https://doi.org/10.1103/PhysRevLett.88.217901} {\bibfield  {journal}
  {\bibinfo  {journal} {Phys. Rev. Lett.}\ }\textbf {\bibinfo {volume} {88}},\
  \bibinfo {pages} {217901} (\bibinfo {year} {2002})}\BibitemShut {NoStop}%
\bibitem [{\citenamefont {Buhrman}\ \emph {et~al.}(2001)\citenamefont
  {Buhrman}, \citenamefont {Cleve}, \citenamefont {Watrous},\ and\
  \citenamefont {de~Wolf}}]{quantumfp}%
  \BibitemOpen
  \bibfield  {author} {\bibinfo {author} {\bibfnamefont {H.}~\bibnamefont
  {Buhrman}}, \bibinfo {author} {\bibfnamefont {R.}~\bibnamefont {Cleve}},
  \bibinfo {author} {\bibfnamefont {J.}~\bibnamefont {Watrous}},\ and\ \bibinfo
  {author} {\bibfnamefont {R.}~\bibnamefont {de~Wolf}},\ }\href
  {https://doi.org/10.1103/PhysRevLett.87.167902} {\bibfield  {journal}
  {\bibinfo  {journal} {Phys. Rev. Lett.}\ }\textbf {\bibinfo {volume} {87}},\
  \bibinfo {pages} {167902} (\bibinfo {year} {2001})}\BibitemShut {NoStop}%
\bibitem [{\citenamefont {Sgobba}\ \emph {et~al.}(2023)\citenamefont {Sgobba},
  \citenamefont {Pallotti}, \citenamefont {Elefante}, \citenamefont
  {Dello~Russo}, \citenamefont {Dequal}, \citenamefont {Siciliani~de Cumis},\
  and\ \citenamefont {Santamaria~Amato}}]{photonics10010072}%
  \BibitemOpen
  \bibfield  {author} {\bibinfo {author} {\bibfnamefont {F.}~\bibnamefont
  {Sgobba}}, \bibinfo {author} {\bibfnamefont {D.~K.}\ \bibnamefont
  {Pallotti}}, \bibinfo {author} {\bibfnamefont {A.}~\bibnamefont {Elefante}},
  \bibinfo {author} {\bibfnamefont {S.}~\bibnamefont {Dello~Russo}}, \bibinfo
  {author} {\bibfnamefont {D.}~\bibnamefont {Dequal}}, \bibinfo {author}
  {\bibfnamefont {M.}~\bibnamefont {Siciliani~de Cumis}},\ and\ \bibinfo
  {author} {\bibfnamefont {L.}~\bibnamefont {Santamaria~Amato}},\ }\href@noop
  {} {\bibfield  {journal} {\bibinfo  {journal} {Photonics}\ }\textbf {\bibinfo
  {volume} {10}} (\bibinfo {year} {2023})}\BibitemShut {NoStop}%
\bibitem [{\citenamefont {Harnchaiwat}\ \emph {et~al.}(2020)\citenamefont
  {Harnchaiwat}, \citenamefont {Zhu}, \citenamefont {Westerberg}, \citenamefont
  {Gauger},\ and\ \citenamefont {Leach}}]{Harnchaiwat:20}%
  \BibitemOpen
  \bibfield  {author} {\bibinfo {author} {\bibfnamefont {N.}~\bibnamefont
  {Harnchaiwat}}, \bibinfo {author} {\bibfnamefont {F.}~\bibnamefont {Zhu}},
  \bibinfo {author} {\bibfnamefont {N.}~\bibnamefont {Westerberg}}, \bibinfo
  {author} {\bibfnamefont {E.}~\bibnamefont {Gauger}},\ and\ \bibinfo {author}
  {\bibfnamefont {J.}~\bibnamefont {Leach}},\ }\href
  {https://doi.org/10.1364/OE.382622} {\bibfield  {journal} {\bibinfo
  {journal} {Opt. Express}\ }\textbf {\bibinfo {volume} {28}},\ \bibinfo
  {pages} {2210} (\bibinfo {year} {2020})}\BibitemShut {NoStop}%
\bibitem [{\citenamefont {Cortes}\ \emph {et~al.}(2022)\citenamefont {Cortes},
  \citenamefont {Lefebvre}, \citenamefont {Lauk}, \citenamefont {Davis},
  \citenamefont {Sinclair}, \citenamefont {Gray},\ and\ \citenamefont
  {Oblak}}]{cortes2022sample}%
  \BibitemOpen
  \bibfield  {author} {\bibinfo {author} {\bibfnamefont {C.~L.}\ \bibnamefont
  {Cortes}}, \bibinfo {author} {\bibfnamefont {P.}~\bibnamefont {Lefebvre}},
  \bibinfo {author} {\bibfnamefont {N.}~\bibnamefont {Lauk}}, \bibinfo {author}
  {\bibfnamefont {M.~J.}\ \bibnamefont {Davis}}, \bibinfo {author}
  {\bibfnamefont {N.}~\bibnamefont {Sinclair}}, \bibinfo {author}
  {\bibfnamefont {S.~K.}\ \bibnamefont {Gray}},\ and\ \bibinfo {author}
  {\bibfnamefont {D.}~\bibnamefont {Oblak}},\ }\href@noop {} {\bibfield
  {journal} {\bibinfo  {journal} {Physical Review Applied}\ }\textbf {\bibinfo
  {volume} {17}},\ \bibinfo {pages} {034067} (\bibinfo {year}
  {2022})}\BibitemShut {NoStop}%
\bibitem [{\citenamefont {Huang}\ \emph {et~al.}(2022)\citenamefont {Huang},
  \citenamefont {Broughton}, \citenamefont {Cotler}, \citenamefont {Chen},
  \citenamefont {Li}, \citenamefont {Mohseni}, \citenamefont {Neven},
  \citenamefont {Babbush}, \citenamefont {Kueng}, \citenamefont {Preskill},\
  and\ \citenamefont {McClean}}]{doi:10.1126/science.abn7293}%
  \BibitemOpen
  \bibfield  {author} {\bibinfo {author} {\bibfnamefont {H.-Y.}\ \bibnamefont
  {Huang}}, \bibinfo {author} {\bibfnamefont {M.}~\bibnamefont {Broughton}},
  \bibinfo {author} {\bibfnamefont {J.}~\bibnamefont {Cotler}}, \bibinfo
  {author} {\bibfnamefont {S.}~\bibnamefont {Chen}}, \bibinfo {author}
  {\bibfnamefont {J.}~\bibnamefont {Li}}, \bibinfo {author} {\bibfnamefont
  {M.}~\bibnamefont {Mohseni}}, \bibinfo {author} {\bibfnamefont
  {H.}~\bibnamefont {Neven}}, \bibinfo {author} {\bibfnamefont
  {R.}~\bibnamefont {Babbush}}, \bibinfo {author} {\bibfnamefont
  {R.}~\bibnamefont {Kueng}}, \bibinfo {author} {\bibfnamefont
  {J.}~\bibnamefont {Preskill}},\ and\ \bibinfo {author} {\bibfnamefont
  {J.~R.}\ \bibnamefont {McClean}},\ }\href@noop {} {\bibfield  {journal}
  {\bibinfo  {journal} {Science}\ }\textbf {\bibinfo {volume} {376}},\ \bibinfo
  {pages} {1182} (\bibinfo {year} {2022})}\BibitemShut {NoStop}%
\bibitem [{\citenamefont {Amaral}\ and\ \citenamefont
  {Tempor{\~a}o}(2019)}]{amaral2019characterization}%
  \BibitemOpen
  \bibfield  {author} {\bibinfo {author} {\bibfnamefont {G.~C.}\ \bibnamefont
  {Amaral}}\ and\ \bibinfo {author} {\bibfnamefont {G.~P.}\ \bibnamefont
  {Tempor{\~a}o}},\ }\href@noop {} {\bibfield  {journal} {\bibinfo  {journal}
  {Quantum Information Processing}\ }\textbf {\bibinfo {volume} {18}},\
  \bibinfo {pages} {1} (\bibinfo {year} {2019})}\BibitemShut {NoStop}%
\bibitem [{\citenamefont {Berglund}(2000)}]{Berglund2000}%
  \BibitemOpen
  \bibfield  {author} {\bibinfo {author} {\bibfnamefont {A.~J.}\ \bibnamefont
  {Berglund}},\ }\href@noop {} {\bibinfo {title} {Quantum coherence and control
  in one- and two-photon optical systems}} (\bibinfo {year} {2000}),\ \Eprint
  {https://arxiv.org/abs/quant-ph/0010001} {arXiv:quant-ph/0010001 [quant-ph]}
  \BibitemShut {NoStop}%
\bibitem [{\citenamefont {Jeffrey}\ \emph {et~al.}(2004)\citenamefont
  {Jeffrey}, \citenamefont {Peters},\ and\ \citenamefont
  {Kwiat}}]{Jeffrey2004}%
  \BibitemOpen
  \bibfield  {author} {\bibinfo {author} {\bibfnamefont {E.}~\bibnamefont
  {Jeffrey}}, \bibinfo {author} {\bibfnamefont {N.~A.}\ \bibnamefont
  {Peters}},\ and\ \bibinfo {author} {\bibfnamefont {P.~G.}\ \bibnamefont
  {Kwiat}},\ }\href {https://doi.org/10.1088/1367-2630/6/1/100} {\bibfield
  {journal} {\bibinfo  {journal} {New Journal of Physics}\ }\textbf {\bibinfo
  {volume} {6}},\ \bibinfo {pages} {100} (\bibinfo {year} {2004})}\BibitemShut
  {NoStop}%
\bibitem [{\citenamefont {Peters}\ \emph {et~al.}(2005)\citenamefont {Peters},
  \citenamefont {Barreiro}, \citenamefont {Goggin}, \citenamefont {Wei},\ and\
  \citenamefont {Kwiat}}]{Peters2005}%
  \BibitemOpen
  \bibfield  {author} {\bibinfo {author} {\bibfnamefont {N.~A.}\ \bibnamefont
  {Peters}}, \bibinfo {author} {\bibfnamefont {J.~T.}\ \bibnamefont
  {Barreiro}}, \bibinfo {author} {\bibfnamefont {M.~E.}\ \bibnamefont
  {Goggin}}, \bibinfo {author} {\bibfnamefont {T.-C.}\ \bibnamefont {Wei}},\
  and\ \bibinfo {author} {\bibfnamefont {P.~G.}\ \bibnamefont {Kwiat}},\ }\href
  {https://doi.org/10.1103/PhysRevLett.94.150502} {\bibfield  {journal}
  {\bibinfo  {journal} {Phys. Rev. Lett.}\ }\textbf {\bibinfo {volume} {94}},\
  \bibinfo {pages} {150502} (\bibinfo {year} {2005})}\BibitemShut {NoStop}%
\bibitem [{\citenamefont {Marcikic}\ \emph {et~al.}(2003)\citenamefont
  {Marcikic}, \citenamefont {de~Riedmatten}, \citenamefont {Tittel},
  \citenamefont {Zbinden},\ and\ \citenamefont {Gisin}}]{Ivan2003}%
  \BibitemOpen
  \bibfield  {author} {\bibinfo {author} {\bibfnamefont {I.}~\bibnamefont
  {Marcikic}}, \bibinfo {author} {\bibfnamefont {H.}~\bibnamefont
  {de~Riedmatten}}, \bibinfo {author} {\bibfnamefont {W.}~\bibnamefont
  {Tittel}}, \bibinfo {author} {\bibfnamefont {H.}~\bibnamefont {Zbinden}},\
  and\ \bibinfo {author} {\bibfnamefont {N.}~\bibnamefont {Gisin}},\ }\href
  {https://doi.org/10.1038/nature01376} {\bibfield  {journal} {\bibinfo
  {journal} {Nature}\ }\textbf {\bibinfo {volume} {421}},\ \bibinfo {pages}
  {509} (\bibinfo {year} {2003})}\BibitemShut {NoStop}%
\bibitem [{\citenamefont {Jing}\ \emph {et~al.}(2019)\citenamefont {Jing},
  \citenamefont {Wang}, \citenamefont {Yu}, \citenamefont {Sun}, \citenamefont
  {Jiang}, \citenamefont {Yang}, \citenamefont {Jiang}, \citenamefont {Luo},
  \citenamefont {Zhang}, \citenamefont {Jiang} \emph
  {et~al.}}]{jing2019entanglement}%
  \BibitemOpen
  \bibfield  {author} {\bibinfo {author} {\bibfnamefont {B.}~\bibnamefont
  {Jing}}, \bibinfo {author} {\bibfnamefont {X.-J.}\ \bibnamefont {Wang}},
  \bibinfo {author} {\bibfnamefont {Y.}~\bibnamefont {Yu}}, \bibinfo {author}
  {\bibfnamefont {P.-F.}\ \bibnamefont {Sun}}, \bibinfo {author} {\bibfnamefont
  {Y.}~\bibnamefont {Jiang}}, \bibinfo {author} {\bibfnamefont {S.-J.}\
  \bibnamefont {Yang}}, \bibinfo {author} {\bibfnamefont {W.-H.}\ \bibnamefont
  {Jiang}}, \bibinfo {author} {\bibfnamefont {X.-Y.}\ \bibnamefont {Luo}},
  \bibinfo {author} {\bibfnamefont {J.}~\bibnamefont {Zhang}}, \bibinfo
  {author} {\bibfnamefont {X.}~\bibnamefont {Jiang}}, \emph {et~al.},\
  }\href@noop {} {\bibfield  {journal} {\bibinfo  {journal} {Nature Photonics}\
  }\textbf {\bibinfo {volume} {13}},\ \bibinfo {pages} {210} (\bibinfo {year}
  {2019})}\BibitemShut {NoStop}%
\bibitem [{\citenamefont {Cramer}\ \emph {et~al.}(2010)\citenamefont {Cramer},
  \citenamefont {Plenio}, \citenamefont {Flammia}, \citenamefont {Somma},
  \citenamefont {Gross}, \citenamefont {Bartlett}, \citenamefont
  {Landon-Cardinal}, \citenamefont {Poulin},\ and\ \citenamefont
  {Liu}}]{cramer2010}%
  \BibitemOpen
  \bibfield  {author} {\bibinfo {author} {\bibfnamefont {M.}~\bibnamefont
  {Cramer}}, \bibinfo {author} {\bibfnamefont {M.~B.}\ \bibnamefont {Plenio}},
  \bibinfo {author} {\bibfnamefont {S.~T.}\ \bibnamefont {Flammia}}, \bibinfo
  {author} {\bibfnamefont {R.}~\bibnamefont {Somma}}, \bibinfo {author}
  {\bibfnamefont {D.}~\bibnamefont {Gross}}, \bibinfo {author} {\bibfnamefont
  {S.~D.}\ \bibnamefont {Bartlett}}, \bibinfo {author} {\bibfnamefont
  {O.}~\bibnamefont {Landon-Cardinal}}, \bibinfo {author} {\bibfnamefont
  {D.}~\bibnamefont {Poulin}},\ and\ \bibinfo {author} {\bibfnamefont {Y.-K.}\
  \bibnamefont {Liu}},\ }\href {https://doi.org/10.1038/ncomms1147} {\bibfield
  {journal} {\bibinfo  {journal} {Nature Communications}\ }\textbf {\bibinfo
  {volume} {1}},\ \bibinfo {pages} {149} (\bibinfo {year} {2010})}\BibitemShut
  {NoStop}%
\bibitem [{\citenamefont {Gisin}\ \emph {et~al.}(2002)\citenamefont {Gisin},
  \citenamefont {Ribordy}, \citenamefont {Tittel},\ and\ \citenamefont
  {Zbinden}}]{Gisin2002}%
  \BibitemOpen
  \bibfield  {author} {\bibinfo {author} {\bibfnamefont {N.}~\bibnamefont
  {Gisin}}, \bibinfo {author} {\bibfnamefont {G.}~\bibnamefont {Ribordy}},
  \bibinfo {author} {\bibfnamefont {W.}~\bibnamefont {Tittel}},\ and\ \bibinfo
  {author} {\bibfnamefont {H.}~\bibnamefont {Zbinden}},\ }\href
  {https://doi.org/10.1103/RevModPhys.74.145} {\bibfield  {journal} {\bibinfo
  {journal} {Rev. Mod. Phys.}\ }\textbf {\bibinfo {volume} {74}},\ \bibinfo
  {pages} {145} (\bibinfo {year} {2002})}\BibitemShut {NoStop}%
\bibitem [{\citenamefont {Nawrath}\ \emph {et~al.}(2023)\citenamefont
  {Nawrath}, \citenamefont {Joos}, \citenamefont {Kolatschek}, \citenamefont
  {Bauer}, \citenamefont {Pruy}, \citenamefont {Hornung}, \citenamefont
  {Fischer}, \citenamefont {Huang}, \citenamefont {Vijayan}, \citenamefont
  {Sittig}, \citenamefont {Jetter}, \citenamefont {Portalupi},\ and\
  \citenamefont {Michler}}]{Nawrath2023}%
  \BibitemOpen
  \bibfield  {author} {\bibinfo {author} {\bibfnamefont {C.}~\bibnamefont
  {Nawrath}}, \bibinfo {author} {\bibfnamefont {R.}~\bibnamefont {Joos}},
  \bibinfo {author} {\bibfnamefont {S.}~\bibnamefont {Kolatschek}}, \bibinfo
  {author} {\bibfnamefont {S.}~\bibnamefont {Bauer}}, \bibinfo {author}
  {\bibfnamefont {P.}~\bibnamefont {Pruy}}, \bibinfo {author} {\bibfnamefont
  {F.}~\bibnamefont {Hornung}}, \bibinfo {author} {\bibfnamefont
  {J.}~\bibnamefont {Fischer}}, \bibinfo {author} {\bibfnamefont
  {J.}~\bibnamefont {Huang}}, \bibinfo {author} {\bibfnamefont
  {P.}~\bibnamefont {Vijayan}}, \bibinfo {author} {\bibfnamefont
  {R.}~\bibnamefont {Sittig}}, \bibinfo {author} {\bibfnamefont
  {M.}~\bibnamefont {Jetter}}, \bibinfo {author} {\bibfnamefont {S.~L.}\
  \bibnamefont {Portalupi}},\ and\ \bibinfo {author} {\bibfnamefont
  {P.}~\bibnamefont {Michler}},\ }\href
  {https://doi.org/https://doi.org/10.1002/qute.202300111} {\bibfield
  {journal} {\bibinfo  {journal} {Advanced Quantum Technologies}\ }\textbf
  {\bibinfo {volume} {6}},\ \bibinfo {pages} {2300111} (\bibinfo {year}
  {2023})}\BibitemShut {NoStop}%
\bibitem [{\citenamefont {Arakawa}\ and\ \citenamefont
  {Holmes}(2020)}]{Arakawa2020}%
  \BibitemOpen
  \bibfield  {author} {\bibinfo {author} {\bibfnamefont {Y.}~\bibnamefont
  {Arakawa}}\ and\ \bibinfo {author} {\bibfnamefont {M.~J.}\ \bibnamefont
  {Holmes}},\ }\href {https://doi.org/10.1063/5.0010193} {\bibfield  {journal}
  {\bibinfo  {journal} {Applied Physics Reviews}\ }\textbf {\bibinfo {volume}
  {7}},\ \bibinfo {pages} {021309} (\bibinfo {year} {2020})}\BibitemShut
  {NoStop}%
\bibitem [{\citenamefont {{Qiskit contributors}}(2023)}]{Qiskit}%
  \BibitemOpen
  \bibfield  {author} {\bibinfo {author} {\bibnamefont {{Qiskit
  contributors}}},\ }\href {https://doi.org/10.5281/zenodo.2573505} {\bibinfo
  {title} {Qiskit: An open-source framework for quantum computing}} (\bibinfo
  {year} {2023})\BibitemShut {NoStop}%
\bibitem [{\citenamefont {Garcia-Escartin}\ and\ \citenamefont
  {Chamorro-Posada}(2013)}]{hom_swaptest}%
  \BibitemOpen
  \bibfield  {author} {\bibinfo {author} {\bibfnamefont {J.~C.}\ \bibnamefont
  {Garcia-Escartin}}\ and\ \bibinfo {author} {\bibfnamefont {P.}~\bibnamefont
  {Chamorro-Posada}},\ }\href {https://doi.org/10.1103/PhysRevA.87.052330}
  {\bibfield  {journal} {\bibinfo  {journal} {Phys. Rev. A}\ }\textbf {\bibinfo
  {volume} {87}},\ \bibinfo {pages} {052330} (\bibinfo {year}
  {2013})}\BibitemShut {NoStop}%
\bibitem [{\citenamefont {Ripper}\ \emph {et~al.}(2023)\citenamefont {Ripper},
  \citenamefont {Amaral},\ and\ \citenamefont {Tempor{\~a}o}}]{swaptest-puc}%
  \BibitemOpen
  \bibfield  {author} {\bibinfo {author} {\bibfnamefont {P.}~\bibnamefont
  {Ripper}}, \bibinfo {author} {\bibfnamefont {G.}~\bibnamefont {Amaral}},\
  and\ \bibinfo {author} {\bibfnamefont {G.}~\bibnamefont {Tempor{\~a}o}},\
  }\href {https://doi.org/10.1007/s11128-023-03961-y} {\bibfield  {journal}
  {\bibinfo  {journal} {Quantum Information Processing}\ }\textbf {\bibinfo
  {volume} {22}},\ \bibinfo {pages} {220} (\bibinfo {year} {2023})}\BibitemShut
  {NoStop}%
\bibitem [{\citenamefont {Foulds}\ \emph {et~al.}(2021)\citenamefont {Foulds},
  \citenamefont {Kendon},\ and\ \citenamefont {Spiller}}]{Foulds_2021}%
  \BibitemOpen
  \bibfield  {author} {\bibinfo {author} {\bibfnamefont {S.}~\bibnamefont
  {Foulds}}, \bibinfo {author} {\bibfnamefont {V.}~\bibnamefont {Kendon}},\
  and\ \bibinfo {author} {\bibfnamefont {T.}~\bibnamefont {Spiller}},\ }\href
  {https://doi.org/10.1088/2058-9565/abe458} {\bibfield  {journal} {\bibinfo
  {journal} {Quantum Science and Technology}\ }\textbf {\bibinfo {volume}
  {6}},\ \bibinfo {pages} {035002} (\bibinfo {year} {2021})}\BibitemShut
  {NoStop}%
\bibitem [{\citenamefont {Calliari}\ \emph {et~al.}(2019)\citenamefont
  {Calliari}, \citenamefont {Tovar}, \citenamefont {Nascimento}, \citenamefont
  {Perlingeiro}, \citenamefont {Amaral},\ and\ \citenamefont
  {Temporao}}]{Calliari:19}%
  \BibitemOpen
  \bibfield  {author} {\bibinfo {author} {\bibfnamefont {F.}~\bibnamefont
  {Calliari}}, \bibinfo {author} {\bibfnamefont {P.}~\bibnamefont {Tovar}},
  \bibinfo {author} {\bibfnamefont {C.}~\bibnamefont {Nascimento}}, \bibinfo
  {author} {\bibfnamefont {B.}~\bibnamefont {Perlingeiro}}, \bibinfo {author}
  {\bibfnamefont {G.}~\bibnamefont {Amaral}},\ and\ \bibinfo {author}
  {\bibfnamefont {G.}~\bibnamefont {Temporao}},\ }\href
  {https://doi.org/10.1364/AO.58.004395} {\bibfield  {journal} {\bibinfo
  {journal} {Appl. Opt.}\ }\textbf {\bibinfo {volume} {58}},\ \bibinfo {pages}
  {4395} (\bibinfo {year} {2019})}\BibitemShut {NoStop}%
\bibitem [{\citenamefont {Ren}\ \emph {et~al.}(2013)\citenamefont {Ren},
  \citenamefont {Du},\ and\ \citenamefont {Deng}}]{PhysRevA.88.012302}%
  \BibitemOpen
  \bibfield  {author} {\bibinfo {author} {\bibfnamefont {B.-C.}\ \bibnamefont
  {Ren}}, \bibinfo {author} {\bibfnamefont {F.-F.}\ \bibnamefont {Du}},\ and\
  \bibinfo {author} {\bibfnamefont {F.-G.}\ \bibnamefont {Deng}},\ }\href
  {https://doi.org/10.1103/PhysRevA.88.012302} {\bibfield  {journal} {\bibinfo
  {journal} {Phys. Rev. A}\ }\textbf {\bibinfo {volume} {88}},\ \bibinfo
  {pages} {012302} (\bibinfo {year} {2013})}\BibitemShut {NoStop}%
\bibitem [{\citenamefont {Khoury}\ and\ \citenamefont
  {Milman}(2011)}]{PhysRevA.83.060301}%
  \BibitemOpen
  \bibfield  {author} {\bibinfo {author} {\bibfnamefont {A.~Z.}\ \bibnamefont
  {Khoury}}\ and\ \bibinfo {author} {\bibfnamefont {P.}~\bibnamefont
  {Milman}},\ }\href {https://doi.org/10.1103/PhysRevA.83.060301} {\bibfield
  {journal} {\bibinfo  {journal} {Phys. Rev. A}\ }\textbf {\bibinfo {volume}
  {83}},\ \bibinfo {pages} {060301(R)} (\bibinfo {year} {2011})}\BibitemShut
  {NoStop}%
\bibitem [{\citenamefont {Lu}\ \emph {et~al.}(2019)\citenamefont {Lu},
  \citenamefont {Lukens}, \citenamefont {Williams}, \citenamefont {Imany},
  \citenamefont {Peters}, \citenamefont {Weiner},\ and\ \citenamefont
  {Lougovski}}]{frequency_bins}%
  \BibitemOpen
  \bibfield  {author} {\bibinfo {author} {\bibfnamefont {H.-H.}\ \bibnamefont
  {Lu}}, \bibinfo {author} {\bibfnamefont {J.~M.}\ \bibnamefont {Lukens}},
  \bibinfo {author} {\bibfnamefont {B.~P.}\ \bibnamefont {Williams}}, \bibinfo
  {author} {\bibfnamefont {P.}~\bibnamefont {Imany}}, \bibinfo {author}
  {\bibfnamefont {N.~A.}\ \bibnamefont {Peters}}, \bibinfo {author}
  {\bibfnamefont {A.~M.}\ \bibnamefont {Weiner}},\ and\ \bibinfo {author}
  {\bibfnamefont {P.}~\bibnamefont {Lougovski}},\ }\href
  {https://doi.org/10.1038/s41534-019-0137-z} {\bibfield  {journal} {\bibinfo
  {journal} {npj Quantum Information}\ }\textbf {\bibinfo {volume} {5}},\
  \bibinfo {pages} {24} (\bibinfo {year} {2019})}\BibitemShut {NoStop}%
\end{thebibliography}%

\appendix
\section{\label{app:a}Protocol for full quantum state tomography}

\begin{figure}[ht]
    \centering
    \includegraphics[width=\linewidth]{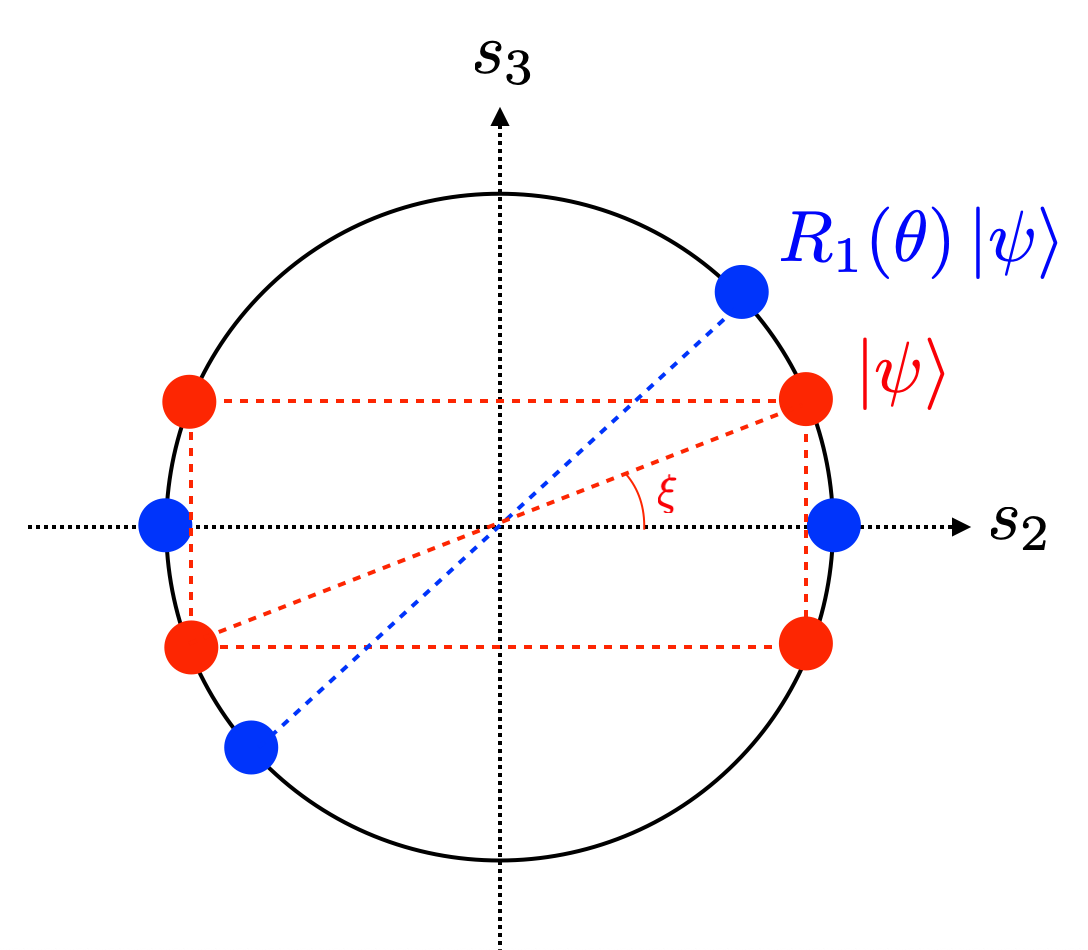}
    \caption{Method for obtaining the signs of Stokes parameters. Originally, the projection of the measured polarization state on the $(s_2,s_3)$ plane could be in any of the four locations shown in red. A small rotation angle $\theta$ around the $s_1$ axis is introduced on both BS inputs and now the four images of the previous possibilities are shown in blue. Depending on the change of the absolute value of $(s_2,s_3)$ one can rule out two of the four possibilities, effectively obtaining the sign of the product $s_2s_3$. Repeating the same reasoning for rotations around one of the other axes, together with the knowledge of the sign of $s_1$ (see text for details), suffices to determine all signs.}
    \label{fig:three}
\end{figure}

It remains to show how one can obtain information on the signs of the Stokes parameters; indeed, the protocol described so far is able to only measure $|s_j|$, as shown in Eq.~\ref{eq:four}. Graphically, this would mean a restriction to the octant $(s_1 >0,s_2>0,s_3>0)$ in Poincaré sphere. There is a way, however, to circumvent this limitation and obtain full information on the Stokes parameters: the first step is increasing the number of measured observables. Instead of only measuring $(I, \sigma_1,\sigma_2,\sigma_3)$ as before, we need to add three more measurements. There are infinitely many solutions, and one of them is depicted in Fig.~\ref{fig:three}. 

Fig.~\ref{fig:three} shows the projection on the $(s_2,s_3)$ plane of the measured polarization state $\ket{\psi}$. As there is no information on the signs of the Stokes parameters, all four red dots are equally likely. Now, we introduce a small rotation angle $\theta$ around the $s_1$ axis, for both BS inputs, in the counter-clockwise direction, resulting in the state $U(\theta)\ket{\psi}$ - which, again, has four possibilities depending on the signs of $(s_2,s_3)$. Now, we compare the measured values of $(s_2,s_3)$ pre- and post-rotation. If $s_2$ decreases and $s_3$ increases, we are in the first or third quadrant, whereas if $s_2$ increases and $s_3$ decreases, we can be sure we are in the second or fourth quadrants. Therefore, we have eliminated two out of four possible cases - in fact, we have determined the sign of the $s_2s_3$ product. The angle $\theta$ should be selected with care, in order to avoid mapping one of the four red points into another red point. For this purpose, the angle $\theta$ should not be fixed, bur rather be a function of the measured Stokes parameters. Let $\xi \equiv \text{tan}^{-1}(|s_3|/|s_2|)$, as depicted in Fig.~\ref{fig:three}. It suffices to select $\theta$ given by:
\begin{eqnarray}
\theta\text{ } && = \frac{\pi}{4}-\frac{\xi}{2}, \xi < \pi/4 \nonumber \\
&& = -\frac{\xi}{2}, \xi \geq \pi/4
\label{eq:angle}
\end{eqnarray}
where the rotation is positive in the counterclockwise direction. Note that $\pi/8 \leq |\theta| \leq \pi/4$. Such a choice for $\theta$ will never result in redundancies, as the rotation will never result in a quadrant change. Then we have:
\begin{equation}
\text{sign}(s_2s_3) = \left\{
\begin{array}{ c l }
    +1 & \quad \textrm{if } |s_2| \textrm{ increases with }\theta  \\
    -1                 & \quad \textrm{otherwise}
  \end{array} \right.
\label{eq:signss2s3}    
\end{equation}

If we repeat the same procedure introducing rotations around the other Poincaré sphere axes, we will obtain the signs of all products $s_1s_2,s_1s_3,s_2s_3$; however, this is still insufficient, as the sign of $s_2s_3$ can be determined from the knowledge of the signs of $s_1s_2$ and $s_1s_3$. In other words, there is a system of three equations and three variables, but one of the equations is linearly dependent of the other two. Indeed, we still have a redundancy of a global factor of $\pm 1$ in the Stokes vector, which means we don't know whether the input polarization state is $\ket{\phi}$ or $\ket{\phi^\perp}$.

For this reason, we introduce a polarization-dependent loss (PDL) in one of the modes, e.g. in front of detector $SPD_0$, as shown in Fig.~\ref{fig:one}. This PDL element should be aligned such that its axes coincide with the horizontal and vertical polarization directions - for example, with a transmission coefficient $\eta_H$ for $\ket{H}$ different from the value $\eta_V$ associated with polarization $\ket{V}$. This addition will change the probability of single clicks on $SPD_0$, depending on whether the incoming photons have positive or negative values of Stokes parameter $s_1$, thus providing the missing information for full state characterization. Note that the actual value of the PDL (i.e., the extinction ratio of the polarizer) plays no role in the calculations of the Stokes parameters. 

In order to illustrate the effect of the PDL, let $C_i$ be the single counts in detector $SPD_i$ and $C_{01}$ be the coincidente counts during the detector's integration time window. Now let $n$ be the number of photon pairs that have impinged on the beamsplitter during the detection time window under consideration. We have:
\begin{eqnarray}
C_0 = && \frac{n}{2}[1-(1-\eta_0)^2]\eta_{PDL} \nonumber \\
C_1 = && \frac{n}{2}[1-(1-\eta_1)^2] \nonumber\\
C_{01} = &&nP_{coinc}\eta_0\eta_1\eta_{PDL}
\label{eq:counts}
\end{eqnarray}
where $P_{coinc}$ is the probability of "bunching" in the BS, which is given by Eq.~\ref{eq:one}, $\eta_{0(1)}$ is the quantum efficiency of $SPD_{0(1)}$ and $\eta_{PDL}$ is the (average) transmission coefficient of the PDL element. Recalling that the PDL has transmissions $\eta_H$ and $\eta_V$ in the horizontal and vertical polarizations, respectively, then:
\begin{equation}
\eta_{PDL} = \eta_H|\braket{\psi | H}|^2 + \eta_V|\braket{\psi | V}|^2
\label{eq:pdl}    
\end{equation}

Note that the horizontal and vertical polarizations can be distinguished by measuring the ratio $C_0/C_1$, i.e., the single counts in the detectors. This provides enough information for determination of the sign of Stokes parameter $s_1$: it suffices to compare $\eta_{PDL}$ with the average value of the horizontal and vertical transmission coefficients:
\begin{equation}
\text{sign}(s_1) = \left\{
\begin{array}{ c l }
    +1 & \quad \textrm{if } \eta_{PDL} \geq (\eta_H+\eta_V)/2 \\
    -1                 & \quad \textrm{otherwise}
  \end{array} \right.
\label{eq:pdl_2}    
\end{equation}
where $\eta_H > \eta_V$ was chosen without loss of generality. Note that in the equality case that the sign is irrelevant as $s_1 = 0$.

In summary, the protocol works in the following way: first off, we measure the absolute values of the Stokes parameters $|s_j|$ by measuring the coincidence rates according to Eq.~\ref{eq:four}, in each case choosing $U = \sigma_j$; then we proceed by calculating two angles $\xi_1 \equiv \text{tan}^{-1}(|s_3|/|s_2|)$ and $\xi_2 \equiv \text{tan}^{-1}(|s_3|/|s_1|)$. Using Eq.~\ref{eq:angle}, we calculate the required angles of rotation around each axis - $R_1(\theta_1)$ and $R_2(\theta_2)$ - and apply the same rotation in both inputs, repeating the measurements of $\sigma_1$ and $\sigma_2$. By calculating the single counts ratio $C_0/C_1$ we have the sign of $s_1$; combining this information with the increase or decrease of the Stokes parameters after the rotations $R_1(\theta_1)$ and $R_2(\theta_2)$, according to Eq.~\ref{eq:signss2s3}, we finally obtain the signs of all Stokes parameters and the protocol is finished.

\section{\label{app:b} Calculation of coincidence rates}

In this appendix we provide calculations for the coincidence rate in Eq. \eqref{eq:one} and the coincidence probability when the input state is given by Fig. 2. 

Let $ a^{\dagger}(\psi) $ and $ b^{\dagger}(\psi) $ be the bosonic creation operators for the two input modes of a symmetrical beam splitter. The BS transformation reads,
\begin{eqnarray}
    a^{\dagger}(\psi) &\rightarrow & \frac{a^{\dagger}(\psi) + i b^{\dagger}(\psi) }{\sqrt{2}} \\
    b^{\dagger}(\psi) &\rightarrow & \frac{b^{\dagger}(\psi) + i a^{\dagger}(\psi) }{\sqrt{2}}
\end{eqnarray}
Modes $ a $ and $ b $ then transform according to,
\begin{eqnarray}
    a^{\dagger}(\psi) b^{\dagger}(\psi') \rightarrow \frac{1}{2}\left(  \underbrace{i  a^{\dagger}(\psi)a^{\dagger}(\psi') + i b^{\dagger}(\psi)b^{\dagger}(\psi')}_\textrm{(i) no coincidences} \right.  \nonumber \\ 
    \left.   \underbrace{a^{\dagger}(\psi)b^{\dagger}(\psi') - b^{\dagger}(\psi) a^{\dagger}(\psi')}_\textrm{(ii) possible coincidences} \right)
    \label{eq:mode_interference}
\end{eqnarray}
\\

\noindent Note the first two terms do not contribute to the coincidence rate, while the remaining ones do. Since,
\begin{eqnarray}
    \vert \psi \rangle = \sqrt{\mathcal{F}} \vert \psi' \rangle + \sqrt{1-\mathcal{F}} e^{i\varphi} \vert \psi'_{\perp} \rangle
\end{eqnarray}
where $ \mathcal{F} = \mathcal{F}(U) $, we may write the same relation in second-quantization notation,
\begin{eqnarray}
    a^{\dagger}(\psi) &=& \sqrt{\mathcal{F}} a^{\dagger}(\psi') + \sqrt{1 - \mathcal{F}} e^{i\varphi} a^{\dagger}(\psi'_{\perp}) \\
    b^{\dagger}(\psi) &=& \sqrt{\mathcal{F}} b^{\dagger}(\psi') + \sqrt{1 - \mathcal{F}} e^{i\varphi} b^{\dagger}(\psi'_{\perp})
\end{eqnarray}
Substituting these relations in Eq. \eqref{eq:mode_interference} we find for the (ii) terms,
\begin{eqnarray}
    \frac{\sqrt{\mathcal{F}}}{2}\left(  \underbrace{  a^{\dagger}(\psi')b^{\dagger}(\psi') -  b^{\dagger}(\psi')a^{\dagger}(\psi')}_\textrm{interference} \right.  \nonumber \\ 
    \left.  \frac{\sqrt{1 - \mathcal{F}}}{2} e^{i\varphi} \underbrace{a^{\dagger}(\psi'_{\perp})b^{\dagger}(\psi') - b^{\dagger}(\psi'_{\perp}) a^{\dagger}(\psi')}_\textrm{no interference} \right)
\end{eqnarray}
Coincidences arise from non-interfering terms, hence we have
\begin{eqnarray}
    P_{coinc} = \vert \pm \frac{1}{2} \sqrt{1 - \mathcal{F}} e^{i\varphi} \vert^{2} \times 2 = \frac{1 - \mathcal{F}}{2}
\end{eqnarray}

Now we proceed to the calculation of the coincidence rate for the internal entanglement case. Consider the interference of two photons, one in state $\ket{\psi}$ and the other in state $(U\otimes I)\ket{\psi}$, where $I$ is the identity operator acting on the time-bin Hilbert space. The state after the BS will be given by:
\begin{equation}
\ket{\psi}_{01}=\left[\frac{i\ket{\psi}_0+\ket{\psi}_1}{\sqrt{2}}\right]\!\otimes\!\left[\frac{(U\otimes I)\left(\ket{\psi}_0+i\ket{\psi}_1\right)}{\sqrt{2}}\right]
\label{eq:six}
\end{equation}
where the subscripts 0 and 1 correspond to the spatial modes where detectors $SPD_0$ and $SPD_1$ are located, respectively. Considering $\ket{\psi} = \alpha\ket{H}\ket{t_0}+\beta\ket{V}\ket{t_1}$ as in Fig.~\ref{fig:two} and replacing in Eq.~\ref{eq:six}, we obtain an expression comprised of 16 terms from all possible combinations of spatial modes, polarization states and time bins, which is given by, up to a normalization constant:

\begin{widetext}
\begin{eqnarray}
\ket{\psi}_{01} && = \text{ }i\alpha^2\ket{H,t_0}_0\ket{U(H),t_0}_0  + i\alpha\beta \ket{H,t_0}_0\ket{U(V),t_1}_0 + i\alpha\beta\ket{V,t_1}_0\ket{U(H),t_0}_0 
+i\beta^2\ket{V,t_1}_0\ket{U(V),t_1}_0
\nonumber \\ && 
-\alpha^2\ket{H,t_0}_0\ket{U(H),t_0}_1 -\alpha\beta \ket{H,t_0}_0\ket{U(V),t_1}_1
-\alpha\beta \ket{V,t_1}_0\ket{U(H),t_0}_1 
-\beta^2\ket{V,t_1}_0\ket{U(V),t_1}_1 
\nonumber \\ &&
+\alpha^2\ket{U(H),t_0}_0\ket{H,t_0}_1 
+\alpha\beta \ket{U(V),t_1}_0\ket{H,t_0}_1 + \alpha\beta \ket{U(H),t_0}_0\ket{V,t_1}_1 + \beta^2\ket{U(V),t_1}_0\ket{V,t_1}_1
\nonumber \\ &&
+ i\alpha^2\ket{H,t_0}_1\ket{U(H),t_0}_1 + i\alpha\beta \ket{H,t_0}_1\ket{U(V),t_1}_1
+ i\alpha\beta \ket{V,t_1}_1\ket{U(H),t_0}_1 + i\beta^2\ket{V,t_1}_1\ket{U(V),t_1}_1
\label{eq:seven}
\end{eqnarray}
\end{widetext}
where the notation $\ket{\psi,t_1} \equiv \ket{\psi}\ket{t_i}$ has been employed for simplicity, and $U(H)/U(V)$ are simplified notations for denoting the image of the unitary operation on the H/V polarization states.

Note that the first four and last four terms in Eq.~\ref{eq:seven} correspond to the cases where both photons are found in the same detector, whereas the remaining eight middle terms correspond to the cases where one photon is found at each detector. In the case where $U = I$, note that the eight middle terms cancel each other out, which is exactly the bunching effect that is expected in such case; we will come back to this result later.\\
In the case where $U = \sigma_1$, which corresponds to the measurement of Stokes parameter $s_1$, we have $U(H) = H$ and $U(V) = -V$, which leads to the cancellation of several terms in Eq.~\ref{eq:seven}, particularly the terms with coefficients $\pm \alpha^2$, $\pm \beta^2$ and $i\alpha\beta$. Therefore, the coincidence probability $P_{coinc}(\sigma_1)$ is given by: 
\begin{equation}
P_{coinc}(\sigma_1) = \frac{4|\alpha|^2|\beta|^2}{2|\alpha|^4+2|\beta|^4+4|\alpha|^2|\beta|^2}
\label{eq:eight}
\end{equation}
Now replacing in Eq.~\ref{eq:four} and using $|\alpha|^2+|\beta|^2 = 1$, we finally obtain:
\begin{equation}
s_1^2 = (2|\alpha|^2-1)^2
\label{eq:nine}
\end{equation}
which coincides with the value obtained by standard quantum state tomography. Indeed, when $\alpha = \beta = 1/\sqrt{2}$, we obtain $s_1$ = 0 and $P_{coinc}(\sigma_1) = 1/2$ as expected, since $\ket{\psi}$ and $\sigma_1\ket{\psi}$ become mutually orthogonal. For any value of $\alpha$, on the other hand, we have $s_2 = s_3 = 0$.

\end{document}